\documentclass[a4paper,11pt]{article}
\usepackage{jheppub} % for details on the use of the package, please see the JINST-author-manual
\usepackage{multirow}
\usepackage{booktabs}
\usepackage{float}
\usepackage{graphicx}
\usepackage{subcaption}
\usepackage{verbatim}
 \arxivnumber{2507.04272} % if you have one

\title{tZ FCNC Case study: LLM Application in signal / Background discrimination analyses in Particle Physics}

\author[a,1]{Saqlain. A,\note{Corresponding author}}
\author[b]{Çabukoğlu. Z,}
\author[c]{Smiesko. J}
\author[a]{and Kartal S}
\affiliation[a]{Istanbul University,\\Istanbul, Türkiye}
\affiliation[b]{Trinity College, Oxford University,\\Oxford, England}
\affiliation[c]{CERN,\\Geneva, Switzerland}

\emailAdd{a.saqlain@ogr.iu.edu.tr} 
\abstract{We present a case study exploring the potential of OpenAI's o3 model as a process-agnostic tool (within fixed topology) for predicting optimized selection cuts in high-energy physics analyses. Specifically, we investigate the effectiveness of the model in separating signal from relevant Standard Model backgrounds in the context of Flavour-Changing Neutral Current (FCNC) top-quark couplings, focusing on the rare decay process $t \rightarrow uZ$. The study is performed at the Future Circular Collider in hadron-hadron mode (FCC-hh) setup. We prompt the o3 model on detector level data for signal and background processes to predict selection cuts that enhance Signal-to-Background (S/B) discrimination. A comparative analysis is then carried out between the efficiencies resulting from o3-predicted cuts, TMVA BDT separation and those derived from traditional manually designed strategies used by a control group; all utilizing the same parameters for the sake of the integrity of the study. Our results demonstrate that the o3 model performs a degree better than the control group, suggesting its promise as a fast and generalizable tool for new physics searches. Meanwhile, BDT results were considerably higher than both the o3 and the traditional cut-based methods. Furthermore, in order to test it's limitations, o3 cuts were applied to data generated via the approved High-Luminosity Large Hadron Collider (HL-LHC) and the proposed High-Energy LHC (HE-LHC) setup in order to examine it's effectiveness at different energy scales when provided with data at FCC energies.}

\begin{document}
\maketitle
\flushbottom

\section{Introduction}
\label{sec:intro}

Although the Standard Model (SM) has been remarkably successful in describing a wide range of particle physics phenomena, it still has certain limitations.~\cite{novaes2000standardmodelintroduction}, physicists often explore the limits of particle physics by proposing and testing various Beyond the Standard Model (BSM) processes. One of these BSM processes is the top‐quark Flavour‐Changing Neutral Current (FCNC) interaction \(t \to uZ\), where t-quark decays to a u quark and a Z boson. This will be the main case for this research. In the SM, this decay is forbidden at tree level and highly suppressed at the loop level due to the Glashow–Iliopoulos–Maiani (GIM) mechanism~\cite{1970PhRvD...2.1285G,aguilar200435,aguilar2009minimal}, with its branching ratio predicted to be of order \(10^{-14}\)~\cite{Agashe:2013}, placing it well below the current LHC reach. However, many BSM models predict \(\mathrm{BR}(t \to qZ)\) to lie in the range \(10^{-4}\)–\(10^{-7}\). Examples include the 2‐Higgs–Doublet Model (2HDM), the quark‐singlet model, the Minimal Supersymmetric Standard Model (MSSM) and its extensions, extended mirror‐fermion models, and warped extra‐dimensions models~\cite{Atwood:1997,Barger_1995,Cao:2007,Yang:1998,Hung:2018,Agashe:2007_warp}. Therefore, the \(tZ\) FCNC process serves as a sensitive probe of BSM physics~\cite{Tait:2000,AguilarSaavedraBranco:2000}. It is to be noted that, to date, no evidence of top‐quark FCNC interactions has been observed, but one must not disregard its discovery potential at the proposed Future Colliders like the High-Luminosity LHC (HL-LHC) with an Integrated Luminosity ($L_{int}$) of 3~ab\(^{-1}\) and a Centre‐of‐Mass energy ($\sqrt{s}$) of 14 TeV~\cite{Aberle:2749422}, High-Energy LHC (HE-LHC) with $L_{int}$ = 15~ab\(^{-1}\) and $\sqrt{s}$ = 27 TeV~\cite{BENEDIKT2018200} and lastly the Future Circular Collider in hadron–hadron mode (FCC‐hh), which will operate at $\sqrt{s}$ = 100 TeV with $L_{int}$ = 30~ab\(^{-1}\)~\cite{ArkaniHamed:2016}.

As with any rare signal, the first step in a detailed study is to isolate it by separating the BSM signal from all relevant SM backgrounds i.e. extract the data from the background noise. Since the 1990s, Machine Learning (ML) techniques have significantly contributed to particle-physics analyses, especially for Signal/Background (S/B) separation. The most common ML practice nowadays utilizes the Toolkit for MultiVariate Analysis (TMVA) package~\cite{Hoecker:2007} within ROOT~\cite{Brun:2019} to train Boosted Decision Trees (BDTs) via AdaBoost. Other frequently used algorithms include \(k\)-nearest neighbours, random forests, and support‐vector machines from the \textsc{scikit‐learn} library~\cite{Pedregosa:2011}, and neural networks implemented via PyTorch~\cite{Paszke:2019}, TensorFlow~\cite{Abadi:2015}, Theano~\cite{AlRfou:2016}, Microsoft Cognitive Toolkit (CNTK)~\cite{SeideAgarwal:2016}, and Keras~\cite{Chollet:2021}. For a more comprehensive review of ML applications in particle physics, see Ref.~\cite{FeickertNachman:2021}.

Since ML models are generally signal‐specific (applicable for only one signal events that they are trained on), there is a rising need to modify them so that they become process-agnostic~\cite{GrossoLetizia:2024} (applicable to multiple signal events) and can be applied across multiple new‐physics scenarios and energies. In this study, we explore the idea of utilizing an LLM model and it's application for a process-agnostic scenario, the model being OpenAI’s latest and most advanced o3 model~\cite{OpenAIo3:2025}. We use Ref.~\cite{LiuMoretti:2021} as a benchmark, adopting the \(ug \to tZ\) process as our signal (case B in~\cite{LiuMoretti:2021}) and treating \(t\bar{t}\), \(t\bar{t}W\), \(t\bar{t}Z\), and \(WZ\) as background events. The main goal is to evaluate o3’s effectiveness and adaptability against first the control study which uses tradition cut-based analyses and secondly, the modern ML techniques specifically the TMVA BDTs. This study may also servse as the first step in evaluating the broader applicability of LLMs in high-energy physics analyses. Recent studies have shown o3’s superiority in competitive‐programming tasks~\cite{OpenAI_CP:2025} as well, making it an attractive prototype for an S/B separation model that is time‐efficient, easily trainable on new datasets, computationally efficient, and accurate in its analyses. We also apply the o3 model's effectiveness when applied on datasets at different energy and detector levels.

This study comprises three main steps. First, we replicate the reference analysis, generating the relevant signal and background samples at FCC-hh energies, and discuss this in detail in Sections.2 and 3. Secondly, we employ the o3 model and the TMVA BDT to predict optimal selection cuts, apply them to our dataset, and compare the results of all three methodologies in terms of effectiveness. This workflow is covered in Section.4 and 5. Lastly, we apply the o3 model (applied on FCC-hh energy) on different detectors and energy levels, in order to evaluate it's effectiveness in those scenarios, in Section.6, and finally we conclude our results in Section.7 at the end.

\section{Theoretical Background}

For this study, we focused on the tZ production which utilized the tZq FCNC (${q=u}$) anomalous couplings. The tZ, after production, further decays into 3 leptons (including an Opposite-sign, Same-Flavor (OSSF) pair), a neutrino and a b-tagged jet as shown in figure \ref{fig:i}.

\begin{figure}[htbp]
  \centering
  \includegraphics[width=.4\textwidth]{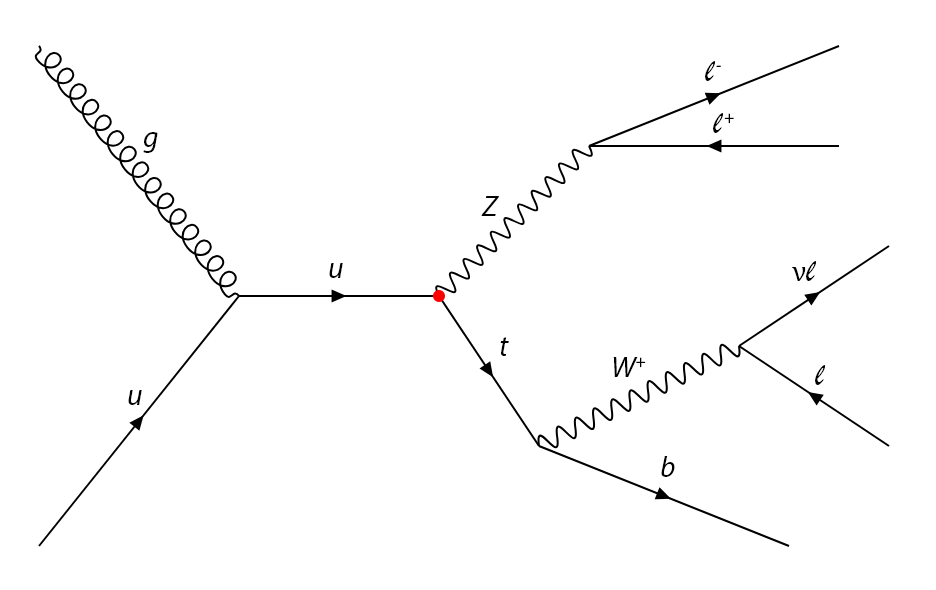}
  \qquad
  \includegraphics[width=.4\textwidth]{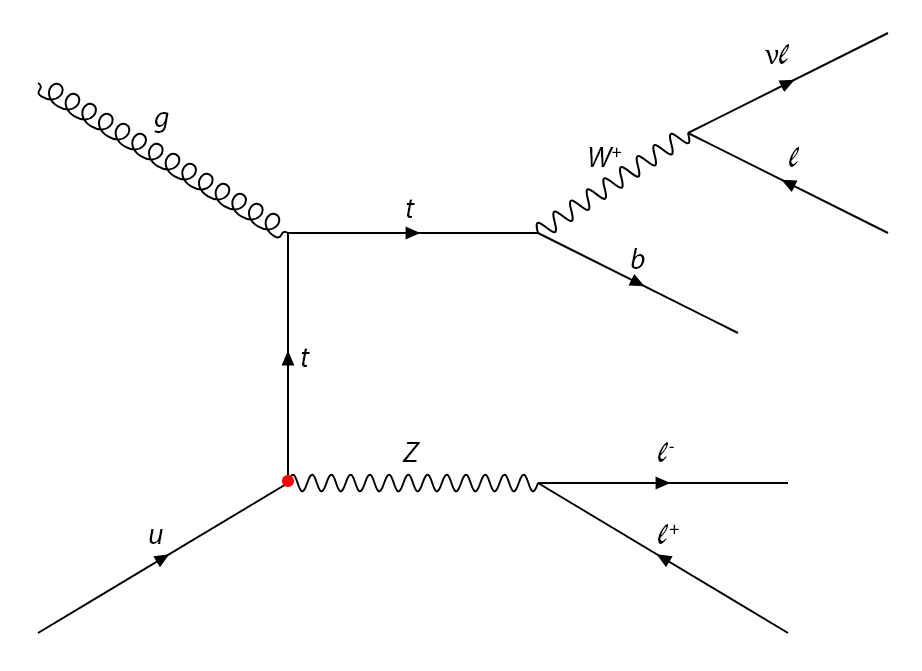}
  \setcounter{figure}{0} 
  \caption{Feynman diagrams representing the \(ug \to tZ\) production and decay via the tZq FCNC anomalous couplings \label{fig:i}}
\end{figure}

The SM model is expanded by using an effective field theory approach which incorporates the tZq ($l{\nu}b{l^+}{l^-}$) FCNC anomalous couplings~\cite{aguilar2009}. The interaction between Z boson, the top quark and an up can be described by the effective Lagrangian as~\cite{aguilar2009}:

\begin{equation} \label{eq:Leff}
\begin{split}
\mathcal{L}_{\text{eff}} 
&= \sum_{q=u,c} \Bigl[ 
    \frac{g}{4c_W m_z}\,\kappa_{tqZ}\,
    \bar{q} \sigma^{\mu\nu \,}
    (\kappa_L P_L + \kappa_R P_R)\,tZ_{\mu\nu}
    \Bigr. \\
&\quad \Bigl. 
    +\,\frac{g}{2c_W}\,\lambda_{tqZ}\,
    \bar{q}\,\gamma^{\mu}\,
    (\lambda_L P_L + \lambda_R P_R) \,tZ_\mu 
\Bigr] \;+\; \text{h.c.}
\end{split}
\end{equation}

here, the $c_W=\cos{\theta_W}$ and ${\theta_W}$ represents the Weinberg angle, $P_{L,R}$ are the left and right-handed chirality projector operators while the effective coupling of the vertices is represented by $\kappa_{tqZ}$ and $\lambda_{tqZ}$. $\kappa_{L,R}$ and $\lambda_{L,R}$ represent the complex chiral parameters and are normalized as $|\kappa_{R}|^2$ + $|\kappa_{L}|^2$ = $|\lambda_{R}|^2$ + $|\lambda_{L}|^2$ = 1, whereas the coupling constant ${g}$ and ${\theta_W}$ govern the Electro-Weak (EW) interaction.

Following the same step as the reference study~\cite{LiuMoretti:2021}, the partial widths for the two tensor variables in the above equation, are calculated by

\begin{equation}
\begin{aligned}
\Gamma\bigl(t \to qZ\bigr)_{\sigma^{\mu\nu}}
&=
\frac{\alpha}{128\,s_W^2\,c_W^2}\,\bigl|\kappa_{tqZ}\bigr|^2\,
\frac{m_t^3}{m_Z^2}
\Bigl[1 - \tfrac{m_Z^2}{m_t^2}\Bigr]^{\!2}
\Bigl[2 + \tfrac{m_Z^2}{m_t^2}\Bigr],
\\[1ex]
\Gamma\bigl(t \to qZ\bigr)_{\gamma^\mu}
&=
\frac{\alpha}{32\,s_W^2\,c_W^2}\,\bigl|\lambda_{tqZ}\bigr|^2\,
\frac{m_t^3}{m_Z^2}
\Bigl[1 - \tfrac{m_Z^2}{m_t^2}\Bigr]^{\!2}
\Bigl[1 + 2\,\tfrac{m_Z^2}{m_t^2}\Bigr].
\end{aligned}
\label{eq:decaywidth_tqZ}
\end{equation}

Agreeing with the SM assumption of dominant top decay to be \(t \to bW^+\)~\cite{li1991qcd} 

\begin{equation}
\Gamma\bigl(t \to bW^+\bigr)
=
\frac{\alpha}{16\,s_W^2}\,\lvert V_{tb}\rvert^2\,
\frac{m_t^3}{m_W^2}
\biggl[\,1 
   - 3\,\frac{m_W^4}{m_t^4} 
   + 2\,\frac{m_W^6}{m_t^6}
\biggr],
\end{equation}

while neglecting the light quark masses, the BR can be calculated as directly dependent on the effective coupling constants~\cite{aguilar200435}:

\begin{equation}
\begin{aligned}
\mathrm{BR}\bigl(t \to qZ\bigr)_{\sigma^{\mu\nu}}
&= 0.172\,\bigl|\kappa_{tqZ}\bigr|^2,\\
\mathrm{BR}\bigl(t \to qZ\bigr)_{\gamma^{\mu}}
&= 0.471\,\bigl|\lambda_{tqZ}\bigr|^2.
\end{aligned}
\label{eq:BR_tqZ}
\end{equation}

In the above calculation, the Next-to-Leading-Order (NLO) QCD correction of top quark decay are also included in a model-independent FCNC framework and the $k$-factor is taken as 1.02~\cite{zhang2009next,drobnak2010flavor}. The relationship between the cross section $\sigma$ and $\kappa_{tqZ}$, $\lambda_{tqZ}$ is highlighted in figure~\ref{fig:fcnc_dependence}

\begin{figure}[ht!]
    \centering

    % First row
    \begin{subfigure}[b]{0.32\textwidth}
        \includegraphics[width=\textwidth]{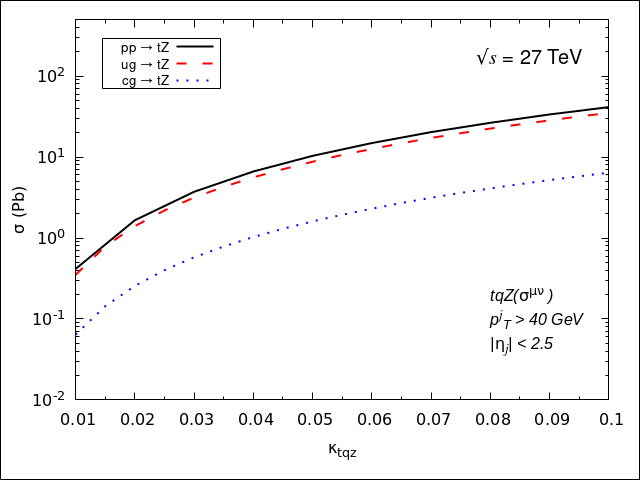}
    \end{subfigure}
    \begin{subfigure}[b]{0.32\textwidth}
        \includegraphics[width=\textwidth]{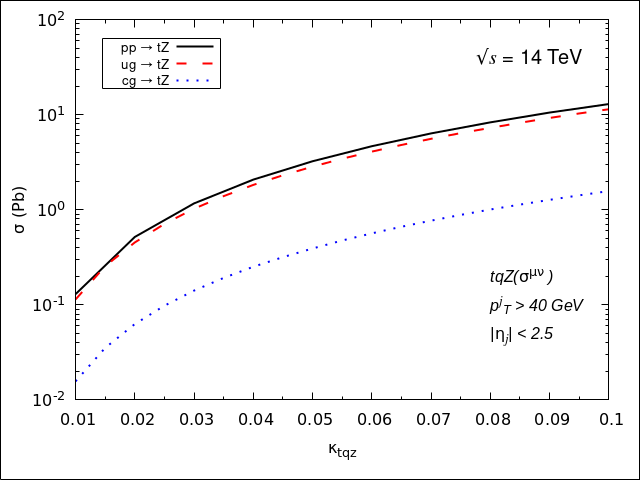}
    \end{subfigure}
    \begin{subfigure}[b]{0.32\textwidth}
        \includegraphics[width=\textwidth]{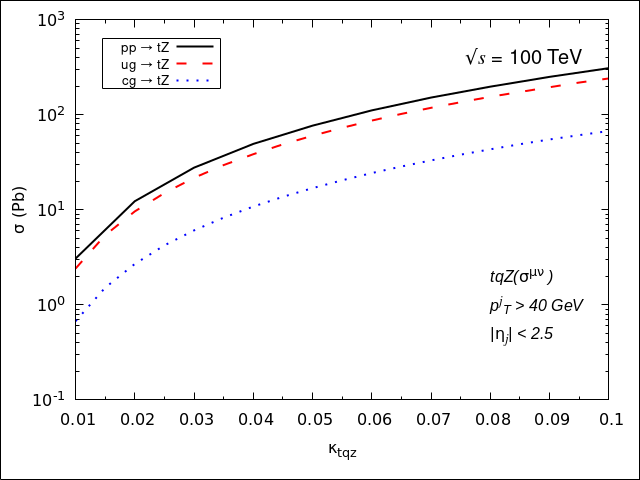}
    \end{subfigure}

    \vspace{0.5em} % optional spacing between rows

    % Second row
    \begin{subfigure}[b]{0.32\textwidth}
        \includegraphics[width=\textwidth]{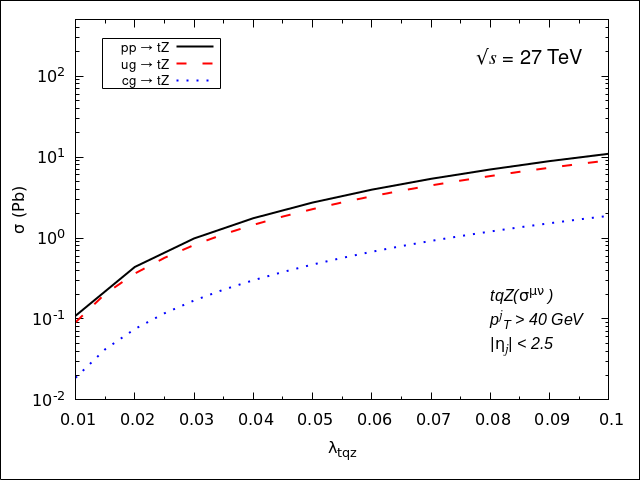}
    \end{subfigure}
    \begin{subfigure}[b]{0.32\textwidth}
        \includegraphics[width=\textwidth]{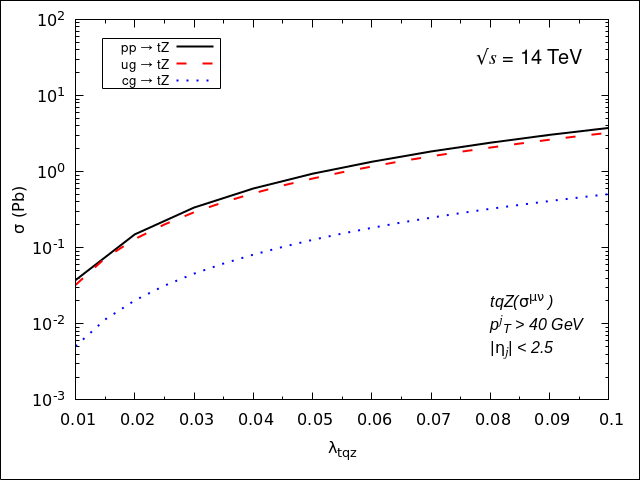}
    \end{subfigure}
    \begin{subfigure}[b]{0.32\textwidth}
        \includegraphics[width=\textwidth]{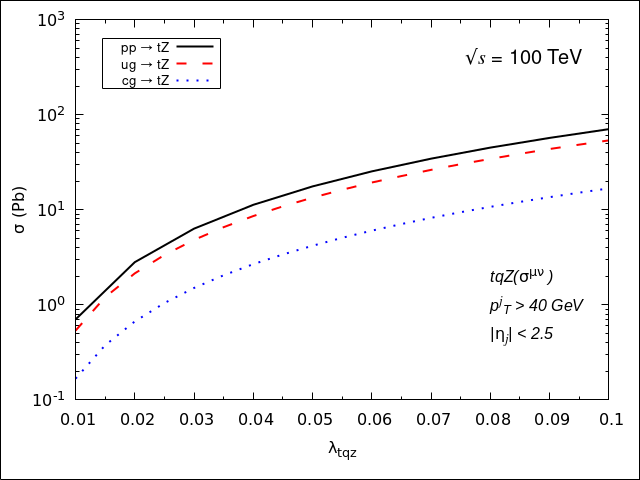}
    \end{subfigure}

    \caption{The variation of the cross section $\sigma$ (in pb) with respect to the FCNC couplings $\kappa_{tqZ}$ (top row) and $\lambda_{tqZ}$ (bottom row) is illustrated for three collider scenarios: HL-LHC (left), HE-LHC (center), and FCC-hh (right). All results are obtained via basic cuts of $p_T^j > 40$ GeV and $|\eta_j| < 2.5.$}
    \label{fig:fcnc_dependence}
\end{figure}

\section{Event generation}
This section deals with the replication of the control study~\cite{LiuMoretti:2021} in order to have the data ready for further comparative analysis. The signal utilized for this study is given as:

\begin{equation}
\begin{aligned}
pp \to t Z, (t \to bW^+ \to bl^+v), (Z \to l^+l^-),
\end{aligned}
\end{equation}

here, $l= \mu, e$. This also includes the charge-conjugate channel $(pp \to \bar{t}Z)$ as well and the signal, overall, is characterized by the decay of 3 leptons , a $b$-jet and Missing Transverse Energy (MET) due to the presence of a neutrino from the decay of $W^{\pm}$. The main backgrounds we are going to be considering are \(t\bar{t}\), \(t\bar{t}W\), \(t\bar{t}Z\), and \(WZ\), all decayed leptonically into electrons and muons.

For the simulation of the signal, we implement the FeynRules package~\cite{alloul2014feynrules} to generate the  Universal FeynRules Output (UFO) files~\cite{degrande2012ufo}. Madgraph5\_aMC@nlo~\cite{alwall2014automated} was employed to generate the signal using the NNPDF23L01 Parton Distribution Functions (PDFs)~\cite{ball2014nnpdf} while taking the renormalization and factorization scales to be $\mu_R=\mu_F=\mu_0/2=(m_t+m_Z)/2$. You can find the value of other parameters below:

\begin{equation}
\begin{aligned}
m_t              &= 173.1\;\mathrm{GeV},  & m_Z              &= 91.1876\;\mathrm{GeV},\\
m_W              &= 80.379\;\mathrm{GeV}, & \alpha_s(m_Z)    &= 0.1181,\\
G_F              &= 1.16637 \times 10^{-5}\;\mathrm{GeV}^{-2}\,. 
\end{aligned}
\end{equation}

Both the signal and background processes, after being generated at LO from Madgraph5\_aMC@nlo, are forwarded to Pythia 8.3~\cite{bierlich2022comprehensive} framework for hadronization. FASTJET 3.2~\cite{cacciari2012fastjet} was utilized for jet clustering with the anti-$k_t$ algorithm with a cone radius of R=0.4~\cite{cacciari2008anti}. Afterwards the events were driven through Delphes 3.5.0~\cite{de2014delphes} to implement detector effects using the default FCC-hh cards and outputted in the form of ROOT files. At the end, MadAnalysis5~\cite{conte2013madanalysis} is used to do the event analysis.

We re-normalize the LO cross sections for the signal to the corresponding higher order QCD results presented in Refs.~\cite{li2011next, zhang2011effective,degrande2015automatic}. Meanwhile, the SM background cross sections are re-normalized to the NLO or Next-NLO (NNLO) order cross sections using Refs.~\cite{lazopoulos2008next,kardos2012top,czakon2013total,https://doi.org/10.23731/cyrm-2017-003,campanario2010nlo,campbell2012t,frederix2018large,frixione2015electroweak}. 

\section{Methodology}

The signal and background event files are obtained from the Madgraph5, Pythia, Delphes pipeline in the form of ROOT files. The ROOT files are converted to .lhco files and provided to the MadAnalysis5 by applying the cuts given by the control group to establish a baseline. 

\begin{figure}[htbp] 
  \centering
  \includegraphics[width=1\textwidth]{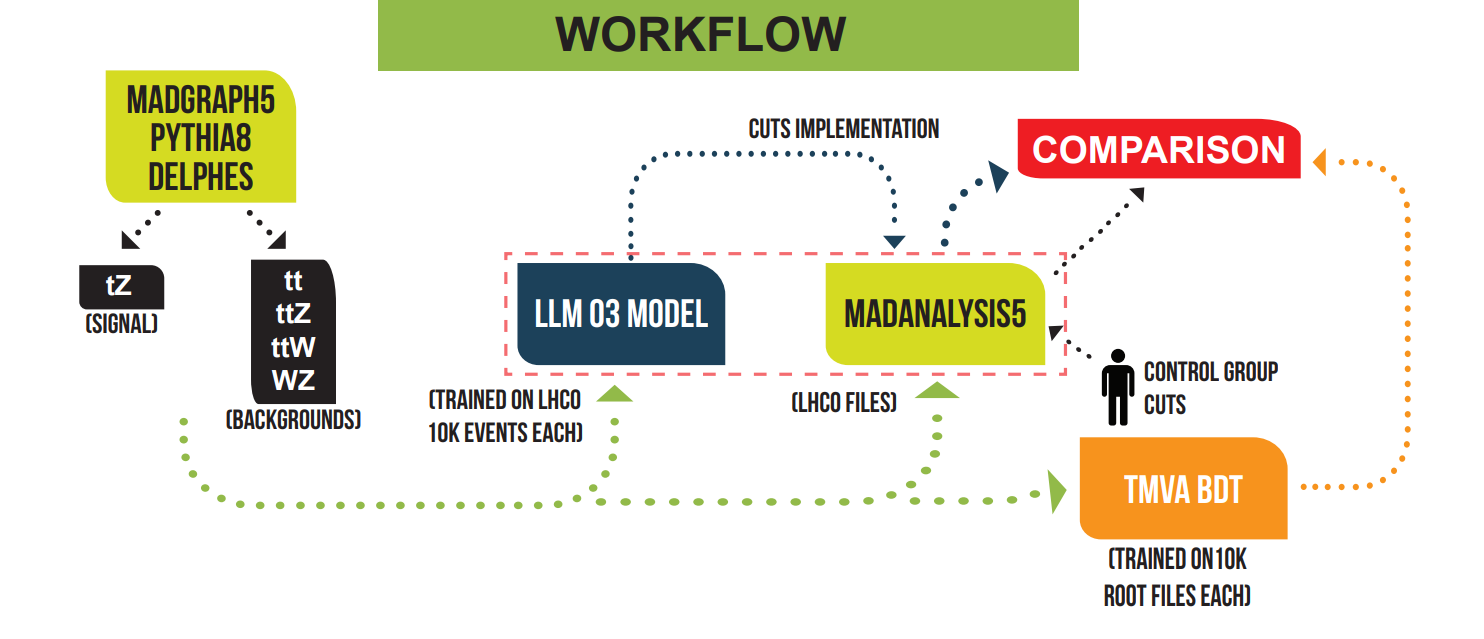}
  \setcounter{figure}{2} 
  \caption{A workflow schematic for this analysis \label{fig:work}}
\end{figure}

Afterwards, the .lhco files are also provided to the LLM o3 model and it is prompted to provide the cuts. These cuts are also applied using MadAnalysis5. Similarly, the outputted ROOT files are trained on by the TMVA BDT and the model is run on the data to obtain the results. Then, all three outputted results are compared using some statistical measures. The schematic workflow is provided in Figure. \ref{fig:work} and more details are expanded on in Section. \ref{Analysis}.

\section{Analysis} \label{Analysis}

This section is divided into four main parts; first, we define the variables and basic cuts that are applied on the dataset to maintain consistency and remove statistical discrepancies. Secondly, we replicate the analysis by utilizing the cuts in the control group study, afterwards, we apply the TMVA BDT on the same data and lastly, the LLM 03 model is provided with the files and cuts are obtained from it.

Based on the control group study, strictly the following variables are utilized for the study (detailed graphs present in Appendix~\ref{app:fig1}) in order to do a more effective comparative analysis:

\begin{itemize}
    \item Number $N_l$ and charge of leptons 
    \item Number of b-jets $N_b$
    \item Transverse momentum ($p_T$) of leptons
    \item $p_T$ of b-jets
    \item Angular distance ($\Delta R$) between the particles
    \item Rapidity ($y(l_1,l_2)$) of the OSSF lepton pair
    \item Transverse mass of W and top quark masses
    \item Invariant mass of the Z boson ($m_{l_1,l_2}$)
\end{itemize}

Moreover, some Basic cuts are applied onto the data set in order to identify objects, defined in the control group article~\cite{LiuMoretti:2021} as:

\begin{equation}
\begin{aligned}
p_T^\ell &> 25~\mathrm{GeV}, & 
p_T^{j/b} &> 30~\mathrm{GeV}, & 
|\eta_i| &< 2.5, \\
\Delta R_{ij} &> 0.4, &  (i, j &= \ell, b, j),
\end{aligned}
\end{equation}

\subsection{Control Group} \label{subsection:control}
All the event ROOT files (signal + background), obtained from Delphes, are converted to Les Houches Collider Output (\texttt{.lhco}) format using root2lhco~\cite{fowlie2016lhcoreadernewcodereading} framework. Do note that even though we will be treating the FCC-hh case as the main case, the other HE-LHC and HL-LHC cases from the control group are also replicated in this subsection.
Below are the cuts defined in the control paper~\cite{LiuMoretti:2021}:

\begin{itemize}
  \item (Cut 1) There are three leptons, among which at least two have positive charge and $P_T > 30$ GeV, and there is exactly one $b$-tagged jet with $P_T > 40$ GeV; the event is rejected if the $P_T$ of the sub-leading jet is greater than 25 GeV. 
  \item (Cut 2) The distance between the OSSF lepton pair should lie within $\Delta R(l_1,l_2) \in [0.2,1.3]$ while the corresponding invariant mass is required to be $|M[l_1,l_2]-m_Z <$ 15 GeV.  
  \item (Cut 3) The transverse masses of the reconstructed $W^{\pm}$ boson and top quark masses are required to satisfy 50 GeV $< M_T^{l_3} <$ 100 GeV and 100 GeV $< M_T^{l_3} <$ 200 GeV, respectively.
  \item (Cut 4) The rapidity of the OSSF lepton pair is required to be $\left| y_{\ell_1 \ell_2} \right| > 1.0$.
\end{itemize}

The effects of the cut selection from the control group are given in the Tables \ref{tab:control_HL_LHC},\ref{tab:control_HE_LHC} and \ref{tab:control_FCC-hh} respectively.

\begin{table}[ht!]
\centering
\caption{Cut flow for the $ug \rightarrow tZ$ signal and background cross sections (in x $10^{-2}$ fb) at HL-LHC with $\kappa_{tqZ} = 0.1$ and $\lambda_{tqZ} = 0.1$, based on the cuts proposed by the control group.}
\renewcommand{\arraystretch}{0.9}
\setlength{\tabcolsep}{5pt}
\small % Shrink font size
\begin{tabular*}{\textwidth}{@{\extracolsep{\fill}}c ccccc}
\toprule
\multirow{2}{*}{\textbf{Cuts}} & \textbf{Signal} & \multicolumn{4}{c}{\textbf{Backgrounds}} \\
                               & $ug \rightarrow tZ$ & $WZ$ & $t\bar{t}$ & $t\bar{t}Z$ & $t\bar{t}W$ \\
\midrule
Basic   & 4221 & 474 & $2.2 \times 10^6$ & 602 & 233 \\
Cut 1   & 272.57   & 23.46  & 1196.97                & 2.83 & 7.12 \\
Cut 2   & 140.35   & 4.51 & 11.22               & 0.57 & 0.071 \\
Cut 3   & 84.895   & 1.13 & 2.83              & 0.11 & 0.019 \\
Cut 4   & 48.55    & 0.24 & 1.21             & 0.034 & 0.009 \\
\bottomrule
\end{tabular*}
\label{tab:control_HL_LHC}
\end{table}

\begin{table}[ht!]
\centering
\caption{Cut flow for the $ug \rightarrow tZ$ signal and background cross sections (in fb) at HE-LHC with $\kappa_{tqZ} = 0.1$ and $\lambda_{tqZ} = 0.1$, based on the cuts proposed by the control group.}
\renewcommand{\arraystretch}{0.9}
\setlength{\tabcolsep}{5pt}
\small % Shrink font size
\begin{tabular*}{\textwidth}{@{\extracolsep{\fill}}c ccccc}
\toprule
\multirow{2}{*}{\textbf{Cuts}} & \textbf{Signal} & \multicolumn{4}{c}{\textbf{Backgrounds}} \\
                               & $ug \rightarrow tZ$ & $WZ$ & $t\bar{t}$ & $t\bar{t}Z$ & $t\bar{t}W$ \\
\midrule
Basic   & 153 & 14.2 & 64628 & 31.6 & 7.7 \\
Cut 1   & 7.47   & 0.476  & 0.823                & 0.06 & 0.089 \\
Cut 2   & 4.1   & 0.07 & 0.017               & 0.014 & 0.0008 \\
Cut 3   & 2.3   & 0.02 & 0.003              & 0.002 & 0.0002 \\
Cut 4   & 1.5    & 0.013 & 0             & 0.0007 & $6.7 \times 10^{-5}$ \\
\bottomrule
\end{tabular*}
\label{tab:control_HE_LHC}
\end{table}

\begin{table}[ht!]
\centering
\caption{Cut flow for the $ug \rightarrow tZ$ signal and background cross sections (in fb) at FCC-hh with $\kappa_{tqZ} = 0.1$ and $\lambda_{tqZ} = 0.1$, based on the cuts proposed by the control group.}
\renewcommand{\arraystretch}{0.9}
\setlength{\tabcolsep}{5pt}
\small % Shrink font size
\begin{tabular*}{\textwidth}{@{\extracolsep{\fill}}c ccccc}
\toprule
\multirow{2}{*}{\textbf{Cuts}} & \textbf{Signal} & \multicolumn{4}{c}{\textbf{Backgrounds}} \\
                               & $ug \rightarrow tZ$ & $WZ$ & $t\bar{t}$ & $t\bar{t}Z$ & $t\bar{t}W$ \\
\midrule
Basic   & 951 & 313 & 697297 & 242 & 43 \\
Cut 1   & 24.1   & 3.77  & 4.98                & 0.33 & 13.38 \\
Cut 2   & 12.57   & 0.49 & 0               & 0.067 & 0.131 \\
Cut 3   & 7.52   & 0.24 & 0              & 0.006 & 0.022 \\
Cut 4   & 5.157    & 0.17 & 0             & 0.003 & 0.017 \\
\bottomrule
\end{tabular*}
\label{tab:control_FCC-hh}
\end{table}

\subsection{TMVA BDT}

In order to employ the TMVA, the variables mentioned above in \ref{subsection:control} are extracted and reconstructed in a ROOT file before providing the files to the TMVA BDT. The following parameters are used:

\begin{itemize}
    \item Number of trees - 600
    \item Gradient boosting
    \item 70/30 - train/test split (same event number as was utilized for the o3 model in \ref{subsection:o3})
    \item Max depth - 3
\end{itemize}

\begin{figure}[htbp]
  \centering
  \includegraphics[width=.4\textwidth]{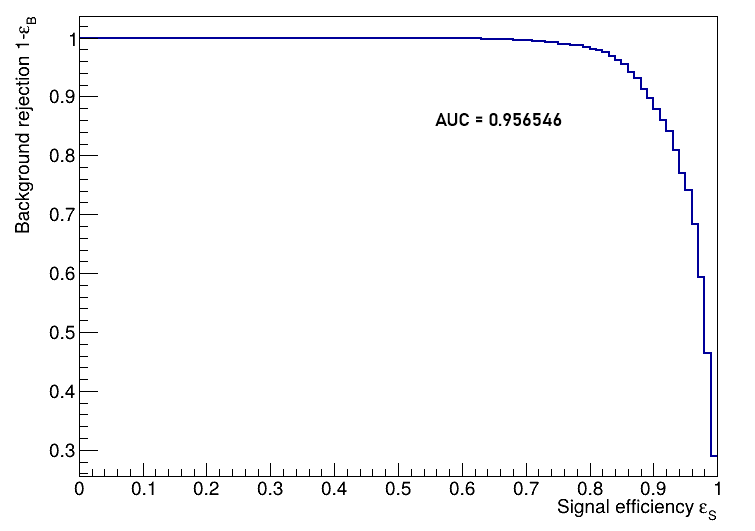}
  \qquad
  \includegraphics[width=.4\textwidth]{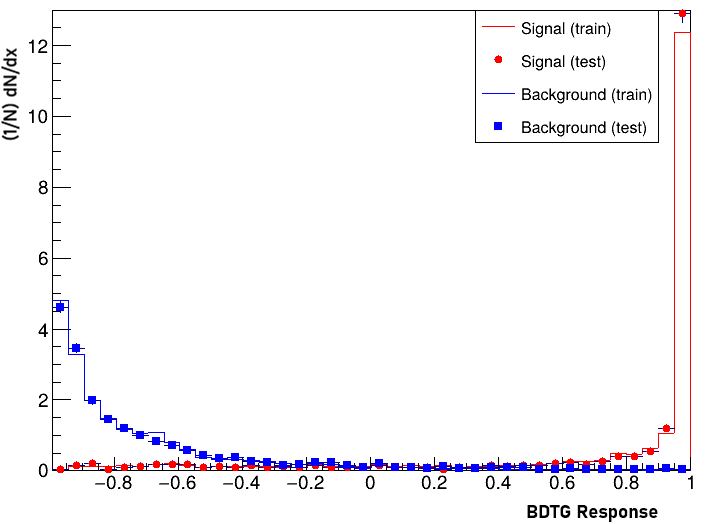}
  \setcounter{figure}{3} 
  \caption{Left: The ROC curve for the BDT model with Background rejection and signal efficiency at its axises. Right: Training error that corresponded to said BDT model \label{fig:train}}
\end{figure}

The top 5 variables that provided the best discrimination factors for the BDT analysis are:

1. $\Delta R$

2. $m_{l_1,l_2}$

3. $N_l$

4. $y(l_1,l_2)$

5. $N_b$

The model was trained iteratively to minimize the training error as much as possible. Both the training graph and the ROC curve for the BDT model can be seen in figure \ref{fig:train}

\subsection{o3 model} \label{subsection:o3}
we directly provided model access to the training data sets in the form of .lhco files by utilizing the project feature on the OpenAI's interface. Afterwards, The model was provided with the following key information:

\begin{itemize}
  \item The label of every \texttt{.lhco} file and what processes are in it.
  \item Information about the \texttt{.lhco} format, syntax and representation of particles inside it.
  \item Full decay chain of each of the processes.
  \item Information about the value of other applicable parameters
\end{itemize}

Then it was prompted to define cuts on each parameter one by one, in separate prompts, with the goal of maximizing signal events while minimizing background ones. The reason for defining the cuts one by one instead of all at once is because it was observed that the model was overwhelmed when asked to define all the cuts at once on all parameters and was not able to go through the data efficiently. This approach mimics how a human analyst might iteratively explore variables to find optimal thresholds. All the prompts and the o3 analysis methodology can be found in appendix~\ref{app:fig}. %reference~\cite{abdullah_bin_saqlain_2025_15733357}.

The o3 model defined the below given cuts for the effective separation of the signal from background events:

\begin{itemize}
  \item (Cut 1) Exactly 3 leptons with two having a positive charge, the requirement for $P_T$ of leading lepton is $P_T^{l1} >$ 200 GeV, of sub-leading lepton is $P_T^{l2} >$ 60 GeV and for the $b$ quarks is $P_T^{ b} >$ 120 GeV. 
  \item (Cut 2) The distance of the OSSF lepton pair should be $\Delta R(l_1,l_2) \in [0.2,1.0]$ and the invariant mass of $Z$ boson is defined as 88 GeV $< |M[l_1,l_2] <$ 95 GeV.
  \item (Cut 3) $M_T$ of top quark is required to be $M_T^{bl_3} <$ 180 GeV and that of $W^{\pm}$ boson is 30 GeV $< M_T^{l_3} <$ 100 GeV.
  \item (Cut 4) The Rapidity of the OSSF pair is required to be $\left| y_{\ell_1 \ell_2} \right| > 2.0$.
\end{itemize}

The resulting cross section cuts effect for the o3 model are given in Table~\ref{tab:o3_FCC-hh}.

\begin{table}[ht!]
\centering
\caption{Cut flow for the $ug \rightarrow tZ$ signal and background cross sections (in fb) at FCC-hh with $\kappa_{tqZ} = 0.1$ and $\lambda_{tqZ} = 0.1$, based on the cuts proposed by o3 model.}
\renewcommand{\arraystretch}{0.9}
\setlength{\tabcolsep}{5pt}
\small % Shrink font size
\begin{tabular*}{\textwidth}{@{\extracolsep{\fill}}c ccccc}
\toprule
\multirow{2}{*}{\textbf{Cuts}} & \textbf{Signal} & \multicolumn{4}{c}{\textbf{Backgrounds}} \\
                               & $ug \rightarrow tZ$ & $WZ$ & $t\bar{t}$ & $t\bar{t}Z$ & $t\bar{t}W$ \\
\midrule
Basic   & 951 & 313 & 697297 & 242 & 43 \\
Cut 1   & 62.81   & 3.603  & 124.18                & 1.65 & 11.32 \\
Cut 2   & 27.53   & 0.898 & 0               & 0.43 & 0.2 \\
Cut 3   & 16.3   & 0 & 0              & 0.07 & 0.006 \\
Cut 4   & 9.06    & 0 & 0            & 0.058 & 0 \\
\bottomrule
\end{tabular*}
\label{tab:o3_FCC-hh}
\end{table}

The o3 model was able to perform high-level reasoning in understanding the event topology, and proposing relevant selection cuts. During the iterative prompting, it adapted its cut suggestions based on histogram feedback and performance summaries, rechecking the variables and efficiencies of signal and backgrounds based on previous iteration of cuts. It was able to parse through the files efficiently, binning the entries and tabulating them after each cut to adjust its strategy, even providing multiple cuts based on the strength of the cut on the background events.

\section{Comparative Analysis}

In this section, we will define the statistical metrics which provide us a better understanding as to the effectiveness of the Analysis done by the o3 and the control group.

A basic signal significance metric, utilizing 

\begin{equation}
\begin{aligned}
Signal \ \  Significance = \frac{S}{\sqrt{S+B}}
\end{aligned}
\end{equation}

where,

$S$ = Number of signal events,

$B$ = Number of background events

Signal efficiency $\epsilon_s$ and background rejection $\epsilon_B$ are two more metrics that provide us with a better understanding of how effective a model is when dealing with S/B discrimination.

\begin{equation}
\begin{aligned}
\epsilon_s = \frac{S_{after\ cut}}{S_{initial}}\ \  ,\ \ \epsilon_B = 1 - \frac{B_{after\ cut}}{B_{initial}}
\end{aligned}
\end{equation}

A comparative analysis based on these values is presented in Table \ref{tab:result}. Moreover, to better evaluate the effectiveness of the o3 model relative to the BDT, the BDT signal significance is reported in the fourth row with respect to the o3 signal efficiency, and in the fifth row with respect to the o3 background rejection.

\begin{table}[ht!]
\centering
\caption{Comparative Analysis of control group (tradition cut-based), BDT and o3 model methodologies based on the given parameters.}
\renewcommand{\arraystretch}{1.2}
\small
\begin{tabular*}{\textwidth}{@{\extracolsep{\fill}}lccc}
\toprule
\textbf{Methods} 
& \textbf{$\epsilon_s$} 
& \textbf{$\epsilon_b$} 
& \textbf{Signal significance} \\
\midrule
Control group   & 0.0054   & 0.9999997   & 12.92 \\
o3 model   & 0.0095   & 0.9999987   & 14.63 \\
BDT   & 0.879  & 0.406   & 26.7 \\
BDT ($\epsilon_s=0.10$)   & 0.10  & 0.99997861   & 31.6 \\
BDT ($\epsilon_b=0.99999$)   & 0.055  & 0.99999   & 23.43 \\
\bottomrule
\end{tabular*}
\label{tab:result}
\end{table}

\begin{comment}

\begin{table}[ht!]
\centering
\caption{Comparative Analysis of the S/B efficiencies for each Analysis at HE-LHC energies.}
\renewcommand{\arraystretch}{1.2}
\small
\begin{tabular*}{\textwidth}{@{\extracolsep{\fill}}lcc}
\toprule
\textbf{Cuts} & \textbf{Efficiency (Control Group)} & \textbf{Efficiency (o3 Model)} \\
\midrule
Basic   & 9.84   & 9.84   \\
Cut 1   & 8.04   & 5.82   \\
Cut 2   & 20.87   &  8.2 \\
Cut 3   & 23.27    & 12.17   \\
Cut 4   & 21.91   & 9.5   \\
\bottomrule
\end{tabular*}
\label{tab:eff_HE_LHC}
\end{table}

\begin{table}[ht!]
\centering
\caption{Comparative Analysis of the S/B efficiencies for each Analysis at FCC-hh energies.}
\renewcommand{\arraystretch}{1.2}
\small
\begin{tabular*}{\textwidth}{@{\extracolsep{\fill}}lcc}
\toprule
\textbf{Cuts} & \textbf{Efficiency (Control Group)} & \textbf{Efficiency (o3 Model)} \\
\midrule
Basic   & 2.73   & 2.73   \\
Cut 1   & 8.64   & 21.07   \\
Cut 2   & 14.39   & 21.87   \\
Cut 3   & 14.38   & 23.55   \\
Cut 4   & 12.54   & 16.88   \\
\bottomrule
\end{tabular*}
\label{tab:eff_FCC_HH}
\end{table}
\end{comment}

It was observed that out of the three methodologies, BDT was the best at preserving signal significance by preserving the most number of signal events but that came at the cost of not rejecting the backgrounds entirely which would cause statistical noise. Comparatively, o3 model provided better signal significance than the control group while still keeping the background events to a minimum. When BDT was evaluated on almost the same extreme metrics $\epsilon_s$ and $\epsilon_b$ as the o3 model (see the fourth and fifth row in Table \ref{tab:result}), it performance dipped exponentially but that is to be expected since we are probing the extreme points of the ROC curve which is not the optimal way to look at a BDT. Another factor in favor of the o3 model is the significantly shorter runtime on the given dataset (approximately 18 minutes), which makes it well suited for time-efficient analysis of large-scale data.

\section{Dependence on Center-of-Mass Energy Scale}
Another problem that the ML models have to deal with is the energy scale difference between datasets at which they are trained at and the experimental dataset, both of which can be of varying energy scales. In this section, we try to expand on the effectiveness of o3 cuts (proposed on dataset generated at energy levels of FCC-hh ($\sqrt{s}=100$ TeV)) at different energy levels; that of HL-LHC ($\sqrt{s}=14$ TeV) and HE-LHC ($\sqrt{s}=27$ TeV). Do note that the particles and processes remain identical but rather the energy scale is changed. The cuts proposed by the o3 model in subsection \ref{subsection:o3} are applied at the data generated by the Madgraph with HL-LHC and HE-LHC parameters. The default HL-LHC and HE-LHC delphes card was utilized for this purpose and their cut-based results are provided in Tables \ref{tab:o3_HL_LHC} and \ref{tab:o3_HE_LHC}.

\begin{table}[H]
\centering
\caption{Cut flow for the $ug \rightarrow tZ$ signal and background cross sections (in x $10^{-2}$ fb) at HL-LHC with $\kappa_{tqZ} = 0.1$ and $\lambda_{tqZ} = 0.1$, based on the cuts proposed by o3 model.}
\renewcommand{\arraystretch}{0.9}
\setlength{\tabcolsep}{5pt}
\small % Shrink font size
\begin{tabular*}{\textwidth}{@{\extracolsep{\fill}}c ccccc}
\toprule
\multirow{2}{*}{\textbf{Cuts}} & \textbf{Signal} & \multicolumn{4}{c}{\textbf{Backgrounds}} \\
                               & $ug \rightarrow tZ$ & $WZ$ & $t\bar{t}$ & $t\bar{t}Z$ & $t\bar{t}W$ \\
\midrule
Basic   & 4221 & 474 & $2.2 \times 10^6$ & 602 & 233 \\
Cut 1   & 172.91   & 1.88  & 207.35                & 4.16 & 1.18 \\
Cut 2   & 83.38   & 0.44 & 0.27               & 1.18 & 0.0009 \\
Cut 3   & 51.96   & 0 & 0              & 0.201 & $5.5 \times 10^{-5}$ \\
Cut 4   & 45.62   & 0 & 0             & 0.199 & $5.5 \times 10^{-5}$ \\
\bottomrule
\end{tabular*}
\label{tab:o3_HL_LHC}
\end{table}

\begin{table}[ht!]
\centering
\caption{Cut flow for the $ug \rightarrow tZ$ signal and background cross sections (in fb) at HE-LHC with $\kappa_{tqZ} = 0.1$ and $\lambda_{tqZ} = 0.1$, based on the cuts proposed by o3 model.}
\renewcommand{\arraystretch}{0.9}
\setlength{\tabcolsep}{5pt}
\small % Shrink font size
\begin{tabular*}{\textwidth}{@{\extracolsep{\fill}}c ccccc}
\toprule
\multirow{2}{*}{\textbf{Cuts}} & \textbf{Signal} & \multicolumn{4}{c}{\textbf{Backgrounds}} \\
                               & $ug \rightarrow tZ$ & $WZ$ & $t\bar{t}$ & $t\bar{t}Z$ & $t\bar{t}W$ \\
\midrule
Basic   & 153 & 14.2 & 64628 & 31.6 & 7.7 \\
Cut 1   & 7.41   & 0.105  & 0.05                & 0.36 & 0.12 \\
Cut 2   & 3.4   & 0.04 & 0               & 0.1 & 0.0002 \\
Cut 3   & 2.02   & 0.001 & 0              & 0.014 & $1.68 \times 10^{-5}$ \\
Cut 4   & 1.53    & 0.001 & 0             & 0.013 & $1.68 \times 10^{-5}$ \\
\bottomrule
\end{tabular*}
\label{tab:o3_HE_LHC}
\end{table}

A comparative analysis in terms of efficiencies between the HL-LHC, HE-LHC and the FCC-hh are provided in table \ref{tab:detectors} below for easier evaluation. 

\begin{table}[ht!]
\centering
\caption{Comparative Analysis of control group (tradition cut-based), BDT and o3 model methodologies based on the given parameters.}
\renewcommand{\arraystretch}{1.2}
\small
\begin{tabular*}{\textwidth}{@{\extracolsep{\fill}}lccc}
\toprule
\textbf{Detectors} 
& \textbf{$\epsilon_s$} 
& \textbf{$\epsilon_b$} 
& \textbf{Signal significance} \\
\midrule
FCC-hh ($\sqrt{s}=100$ Tev)   & 0.0095   & 0.99999353   & 14.63 \\
HE-LHC ($\sqrt{s}=27$ Tev)   & 0.011   & 0.99998638   & 14.32 \\
HL-LHC ($\sqrt{s}=14$ Tev)   & 0.0108  & 0.99999397   & 18.153 \\
\bottomrule
\end{tabular*}
\label{tab:detectors}
\end{table}

From the table, we can see that the efficiencies of the signal and background remain somewhat constant or better (in the case of HL-LHC). This indicates the effectiveness of o3 in selection cuts and can be expanded upon further in future studies. Do note that this table deals with the exact numerical number of events rather than the cross-sections normalized to the relevant energies (given in Tables~\ref{tab:o3_HL_LHC} and~\ref{tab:o3_HE_LHC}).

\section{Conclusions}

In this study, we analyzed the efficiency of cuts proposed by the o3 model with respect to those proposed by the control group on the FCNC  $t \to uZ$ anomalous couplings for the signal at FCC-hh energies. We provided the o3 model and the BDT with a set of 50k events in total, in \texttt{.lhco} format, and then used it to predict selection cuts. We performed a comparative analysis of the efficiency of the signal and relevant SM background cuts from the control group, the o3 model and the BDT. In just under 20 minutes of analysis, the o3 model was able to provide cuts that were competitive with the control group cut selections. The BDT resulted in better signal significance overall which conforms with it being a superior technique than cut-based analysis. Lastly, at different energy levels, it was seen that the o3 has the same problem like other models which are trained at one energy i.e. they are unable to perform better or comparable at other energy levels. This does not diminish the significant fact that o3, despite not being a model specifically designed for particle physics, was able to perform accurate signal separation and may serve as a precursor to a new generation of process-agnostic tool (within fixed topology) for future analyses.

%\appendix

\acknowledgments

The numerical calculations reported in this paper were fully/partially performed at TUBITAK ULAKBIM, High Performance and Grid Computing Center (TRUBA resources).

% Bibliography

%% [A] Recommended: using JHEP.bst file
\bibliographystyle{JHEP}
\bibliography{biblio.bib}

@misc{novaes2000standardmodelintroduction,
      title={Standard Model: An Introduction}, 
      author={S. F. Novaes},
      year={2000},
      eprint={hep-ph/0001283},
      archivePrefix={arXiv},
      primaryClass={hep-ph},
      url={https://arxiv.org/abs/hep-ph/0001283}, 
}

@ARTICLE{1970PhRvD...2.1285G,
       author = {{Glashow}, S.~L. and {Iliopoulos}, J. and {Maiani}, L.},
        title = "{Weak Interactions with Lepton-Hadron Symmetry}",
      journal = {\prd},
         year = 1970,
        month = oct,
       volume = {2},
       number = {7},
        pages = {1285-1292},
          doi = {10.1103/PhysRevD.2.1285},
       adsurl = {https://ui.adsabs.harvard.edu/abs/1970PhRvD...2.1285G}
}

@misc{aguilar200435,
  title={35.2695 A. vol. 35},
  author={Aguilar-Saavedra, JA2004AcPPB},
  journal={Acta Phys. Polon. B},
  pages={2695},
  year={2004}
}

@article{aguilar2009minimal,
  title={A minimal set of top-Higgs anomalous couplings},
  author={Aguilar-Saavedra, Juan Antonio},
  journal={Nuclear physics B},
  volume={821},
  number={1-2},
  pages={215--227},
  year={2009},
  publisher={Elsevier}
}

@misc{Agashe:2013,
      title={Snowmass 2013 Top quark working group report}, 
      author={K. Agashe and R. Erbacher and C. E. Gerber and K. Melnikov and R. Schwienhorst and A. Mitov and M. Vos and S. Wimpenny and J. Adelman and M. Baumgart and A. Garcia-Bellido and A. Loginov and A. Jung and M. Schulze and J. Shelton and N. Craig and M. Velasco and T. Golling and J. Hubisz and A. Ivanov and M. Perelstein and S. Chekanov and J. Dolen and J. Pilot and R. Pöschl and B. Tweedie and S. Alioli and B. Alvarez-Gonzalez and D. Amidei and T. Andeen and A. Arce and B. Auerbach and A. Avetisyan and M. Backovic and Y. Bai and M. Begel and S. Berge and C. Bernard and C. Bernius and S. Bhattacharya and K. Black and A. Blondel and K. Bloom and T. Bose and J. Boudreau and J. Brau and A. Broggio and G. Brooijmans and E. Brost and R. Calkins and D. Chakraborty and T. Childress and G. Choudalakis and V. Coco and J. S. Conway and C. Degrande and A. Delannoy and F. Deliot and L. Dell'Asta and E. Drueke and B. Dutta and A. Effron and K. Ellis and J. Erdmann and J. Evans and C. Feng and E. Feng and A. Ferroglia and K. Finelli and W. Flanagan and I. Fleck and A. Freitas and F. Garberson and R. Gonzalez Suarez and M. L. Graesser and N. Graf and Z. Greenwood and J. George and C. Group and A. Gurrola and G. Hammad and T. Han and Z. Han and U. Heintz and S. Hoeche and T. Horiguchi and I. Iashvili and A. Ismail and S. Jain and P. Janot and W. Johns and J. Joshi and A. Juste and T. Kamon and C. Kao and Y. Kats and A. Katz and M. Kaur and R. Kehoe and W. Keung and S. Khalil and A. Khanov and A. Kharchilava and N. Kidonakis and C. Kilic and N. Kolev and A. Kotwal and J. Kraus and D. Krohn and M. Kruse and A. Kumar and S. Lee and E. Luiggi and S. Mantry and A. Melo and D. Miller and G. Moortgat-Pick and M. Narain and N. Odell and Y. Oksuzian and M. Oreglia and A. Penin and Y. Peters and C. Pollard and S. Poss and H. B. Prosper S. Rappoccio and S. Redford and M. Reece and F. Rizatdinova and P. Roloff and R. Ruiz and M. Saleem and B. Schoenrock and C. Schwanenberger and T. Schwarz and K. Seidel and E. Shabalina and P. Sheldon and F. Simon and K. Sinha and P. Skands and P. Skubik and G. Sterman and D. Stolarski and J. Strube and J. Stupak and S. Su and M. Tesar and S. Thomas and E. Thompson and P. Tipton and E. Varnes and N. Vignaroli and J. Virzi and M. Vogel and D. Walker and K. Wang and B. Webber and J. D. Wells and S. Westhoff and D. Whiteson and M. Williams and S. Wu and U. Yang and H. Yokoya and H. Yoo and H. Zhang and N. Zhou and H. Zhu and J. Zupan},
      year={2013},
      eprint={1311.2028},
      archivePrefix={arXiv},
      primaryClass={hep-ph},
      url={https://arxiv.org/abs/1311.2028}, 
}

@article{Atwood:1997,
  title = {Phenomenology of two Higgs doublet models with flavor-changing neutral currents},
  author = {Atwood, David and Reina, Laura and Soni, Amarjit},
  journal = {Phys. Rev. D},
  volume = {55},
  issue = {5},
  pages = {3156--3176},
  numpages = {0},
  year = {1997},
  month = {Mar},
  publisher = {American Physical Society},
  doi = {10.1103/PhysRevD.55.3156},
  url = {https://link.aps.org/doi/10.1103/PhysRevD.55.3156}
}

@article{Barger_1995,
   title={Quark singlets: Implications and constraints},
   volume={52},
   ISSN={0556-2821},
   url={http://dx.doi.org/10.1103/PhysRevD.52.1663},
   DOI={10.1103/physrevd.52.1663},
   number={3},
   journal={Physical Review D},
   publisher={American Physical Society (APS)},
   author={Barger, V. and Berger, M. S. and Phillips, R. J. N.},
   year={1995},
   month=aug, pages={1663–1683} }

@article{Cao:2007,
  title = {Supersymmetry-induced flavor-changing neutral-current top-quark processes at the CERN Large Hadron Collider},
  author = {Cao, J. J. and Eilam, G. and Frank, M. and Hikasa, K. and Liu, G. L. and Turan, I. and Yang, J. M.},
  journal = {Phys. Rev. D},
  volume = {75},
  issue = {7},
  pages = {075021},
  numpages = {22},
  year = {2007},
  month = {Apr},
  publisher = {American Physical Society},
  doi = {10.1103/PhysRevD.75.075021},
  url = {https://link.aps.org/doi/10.1103/PhysRevD.75.075021}
}

@article{Yang:1998,
  title = {Flavor-changing top quark decays in R-parity-violating supersymmetric models},
  author = {Yang, Jin Min and Young, Bing-Lin and Zhang, X.},
  journal = {Phys. Rev. D},
  volume = {58},
  issue = {5},
  pages = {055001},
  numpages = {6},
  year = {1998},
  month = {Jul},
  publisher = {American Physical Society},
  doi = {10.1103/PhysRevD.58.055001},
  url = {https://link.aps.org/doi/10.1103/PhysRevD.58.055001}
}

@article{Hung:2018,
title = {Top quark rare decays via loop-induced FCNC interactions in extended mirror fermion model},
journal = {Nuclear Physics B},
volume = {927},
pages = {166-183},
year = {2018},
issn = {0550-3213},
doi = {https://doi.org/10.1016/j.nuclphysb.2017.12.014},
url = {https://www.sciencedirect.com/science/article/pii/S0550321317303991},
author = {P.Q. Hung and Yu-Xiang Lin and Chrisna Setyo Nugroho and Tzu-Chiang Yuan}
}

@article{Agashe:2007_warp,
  title={Collider signals of top quark flavor violation from a warped extra dimension},
  author={Agashe, Kaustubh and Perez, Gilad and Soni, Amarjit},
  journal={Physical Review D—Particles, Fields, Gravitation, and Cosmology},
  volume={75},
  number={1},
  pages={015002},
  year={2007},
  publisher={APS}
}

@article{Tait:2000,
  title={Single top quark production as a window to physics beyond the standard model},
  author={Tait, Tim MP and Yuan, C-P},
  journal={Physical Review D},
  volume={63},
  number={1},
  pages={014018},
  year={2000},
  publisher={APS}
}

@article{AguilarSaavedraBranco:2000,
  title={Probing top flavour-changing neutral scalar couplings at the CERN LHC},
  author={Aguilar-Saavedra, JA and Branco, Gustavo Castello},
  journal={Physics Letters B},
  volume={495},
  number={3-4},
  pages={347--356},
  year={2000},
  publisher={Elsevier}
}

@book{Aberle:2749422,
      author        = "Aberle, O. and Béjar Alonso, I and Brüning, O and
                       Fessia, P and Rossi, L and Tavian, L and Zerlauth, M and
                       Adorisio, C. and Adraktas, A. and Ady, M. and Albertone, J.
                       and Alberty, L. and Alcaide Leon, M. and Alekou, A. and
                       Alesini, D. and Ferreira, B. Almeida and Lopez, P. Alvarez
                       and Ambrosio, G. and Andreu Munoz, P. and Anerella, M. and
                       Angal-Kalinin, D. and Antoniou, F. and Apollinari, G. and
                       Apollonio, A. and Appleby, R. and Arduini, G. and Alonso,
                       B. Arias and Artoos, K. and Atieh, S. and Auchmann, B. and
                       Badin, V. and Baer, T. and Baffari, D. and Baglin, V. and
                       Bajko, M. and Ball, A. and Ballarino, A. and Bally, S. and
                       Bampton, T. and Banfi, D. and Barlow, R. and Barnes, M. and
                       Barranco, J. and Barthelemy, L. and Bartmann, W. and
                       Bartosik, H. and Barzi, E. and Battistin, M. and
                       Baudrenghien, P. and Alonso, I. Bejar and Belomestnykh, S.
                       and Benoit, A. and Ben-Zvi, I. and Bertarelli, A. and
                       Bertolasi, S. and Bertone, C. and Bertran, B. and Bestmann,
                       P. and Biancacci, N. and Bignami, A. and Bliss, N. and
                       Boccard, C. and Body, Y. and Borburgh, J. and Bordini, B.
                       and Borralho, F. and Bossert, R. and Bottura, L. and
                       Boucherie, A. and Bozzi, R. and Bracco, C. and Bravin, E.
                       and Bregliozzi, G. and Brett, D. and Broche, A. and
                       Brodzinski, K. and Broggi, F. and Bruce, R. and Brugger, M.
                       and Brüning, O. and Buffat, X. and Burkhardt, H. and
                       Burnet, J. and Burov, A. and Burt, G. and Cabezas, R. and
                       Cai, Y. and Calaga, R. and Calatroni, S. and Capatina, O.
                       and Capelli, T. and Cardon, P. and Carlier, E. and Carra,
                       F. and Carvalho, A. and Carver, L.R. and Caspers, F. and
                       Cattenoz, G. and Cerutti, F. and Chancé, A. and Rodrigues,
                       M. Chastre and Chemli, S. and Cheng, D. and Chiggiato, P.
                       and Chlachidze, G. and Claudet, S. and Coello De Portugal,
                       JM. and Collazos, C. and Corso, J. and Costa Machado, S.
                       and Costa Pinto, P. and Coulinge, E. and Crouch, M. and
                       Cruikshank, P. and Cruz Alaniz, E. and Czech, M. and
                       Dahlerup-Petersen, K. and Dalena, B. and Daniluk, G. and
                       Danzeca, S. and Day, H. and De Carvalho Saraiva, J. and De
                       Luca, D. and De Maria, R. and De Rijk, G. and De Silva, S.
                       and Dehning, B. and Delayen, J. and Deliege, Q. and
                       Delille, B. and Delsaux, F. and Denz, R. and Devred, A. and
                       Dexter, A. and Di Girolamo, B. and Dietderich, D. and
                       Dilly, J.W. and Doherty, A. and Dos Santos, N. and Drago,
                       A. and D.Drskovic and Ramos, D. Duarte and Ducimetière, L.
                       and Efthymiopoulos, I. and Einsweiler, K. and Esposito, L.
                       and Esteban Muller, J. and Evrard, S. and Fabbricatore, P.
                       and Farinon, S. and Fartoukh, S. and Faus-Golfe, A. and
                       Favre, G. and Felice, H. and Feral, B. and Ferlin, G. and
                       Ferracin, P. and Ferrari, A. and Ferreira, L. and Fessia,
                       P. and Ficcadenti, L. and Fiotakis, S. and Fiscarelli, L.
                       and Fitterer, M. and Fleiter, J. and Foffano, G. and Fol,
                       E. and Folch, R. and Foraz, K. and Foussat, A. and Frankl,
                       M. and Frasciello, O. and Fraser, M. and Menendez, P.
                       Freijedo and Fuchs, J-F. and Furuseth, S. and Gaddi, A. and
                       Gallilee, M. and Gallo, A. and Alia, R. Garcia and Gavela,
                       H. Garcia and Matos, J. Garcia and Garcia Morales, H. and
                       Valdivieso, A. Garcia-Tabares and Garino, C. and Garion, C.
                       and Gascon, J. and Gasnier, Ch. and Gentini, L. and
                       Gentsos, C. and Ghosh, A. and Giacomel, L. and Hernandez,
                       K. Gibran and Gibson, S. and Ginburg, C. and Giordano, F.
                       and Giovannozzi, M. and Goddard, B. and Gomes, P. and
                       Gonzalez De La Aleja Cabana, M. and Goudket, P. and
                       Gousiou, E. and Gradassi, P. and Costa, A. Granadeiro and
                       Grand-Clément, L. and Grillot, S. and Guillaume, JC. and
                       Guinchard, M. and Hagen, P. and Hakulinen, T. and Hall, B.
                       and Hansen, J. and Heredia Garcia, N. and Herr, W. and
                       Herty, A. and Hill, C. and Hofer, M. and Höfle, W. and
                       Holzer, B. and Hopkins, S. and Hrivnak, J. and Iadarola, G.
                       and Infantino, A. and Bermudez, S. Izquierdo and Jakobsen,
                       S. and Jebramcik, M.A. and Jenninger, B. and Jensen, E. and
                       Jones, M. and Jones, R. and Jones, T. and Jowett, J. and
                       Juchno, M. and Julie, C. and Junginger, T. and Kain, V. and
                       Kaltchev, D. and Karastathis, N. and Kardasopoulos, P. and
                       Karppinen, M. and Keintzel, J. and Kersevan, R. and
                       Killing, F. and Kirby, G. and Korostelev, M. and Kos, N.
                       and Kostoglou, S. and Kozsar, I. and Krasnov, A. and Krave,
                       S. and Krzempek, L. and Kuder, N. and Kurtulus, A. and
                       Kwee-Hinzmann, R. and Lackner, F. and Lamont, M. and
                       Lamure, A.L. and m, L. Lari and Lazzaroni, M. and Le
                       Garrec, M. and Lechner, A. and Lefevre, T. and Leuxe, R.
                       and Li, K. and Li, Z. and Lindner, R. and Lindstrom, B. and
                       Lingwood, C. and Löffler, C. and Lopez, C. and
                       Lopez-Hernandez, LA. and Losito, R. and Maciariello, F. and
                       Macintosh, P. and Maclean, E.H. and Macpherson, A. and
                       Maesen, P. and Magnier, C. and Durand, H. Mainaud and
                       Malina, L. and Manfredi, M. and Marcellini, F. and
                       Marchevsky, M. and Maridor, S. and Marinaro, G. and
                       Marinov, K. and Markiewicz, T. and Marsili, A. and Martinez
                       Urioz, P. and Martino, M. and Masi, A. and Mastoridis, T.
                       and Mattelaer, P. and May, A. and Mazet, J. and Mcilwraith,
                       S. and McIntosh, E. and Medina Medrano, L. and Mejica
                       Rodriguez, A. and Mendes, M. and Menendez, P. and Mensi, M.
                       and Mereghetti, A. and Mergelkuhl, D. and Mertens, T. and
                       Mether, L. and Métral, E. and Migliorati, M. and Milanese,
                       A. and Minginette, P. and Missiaen, D. and Mitsuhashi, T.
                       and Modena, M. and Mokhov, N. and Molson, J. and Monneret,
                       E. and Montesinos, E. and Moron-Ballester, R. and Morrone,
                       M. and Mostacci, A. and Mounet, N. and Moyret, P. and
                       Muffat, P. and Muratori, B. and Muttoni, Y. and Nakamoto,
                       T. and Navarro-Tapia, M. and Neupert, H. and Nevay, L. and
                       Nicol, T. and Nilsson, E. and Ninin, P. and Nobrega, A. and
                       Noels, C. and Nolan, E. and Nosochkov, Y. and Nuiry, FX.
                       and Oberli, L. and Ogitsu, T. and Ohmi, K. and Olave R. and
                       Oliveira, J. and Orlandi, Ph. and Ortega, P. and Osborne,
                       J. and Otto, T. and Palumbo, L. and Papadopoulou, S. and
                       Papaphilippou, Y. and Paraschou, K. and Parente, C. and
                       Paret, S. and Park, H. and Parma, V. and Pasquino, Ch. and
                       Patapenka, A. and Patnaik, L. and Pattalwar, S. and Payet,
                       J. and Pechaud, G. and Pellegrini, D. and Pepinster, P. and
                       Perez, J. and Espinos, J. Perez and Marcone, A. Perillo and
                       Perin, A. and Perini, P. and Persson, T.H.B. and Peterson,
                       T. and Pieloni, T. and Pigny, G. and Pinheiro de Sousa,
                       J.P. and Pirotte, O. and Plassard, F. and Pojer, M. and
                       Pontercorvo, L. and Poyet, A. and Prelipcean, D. and Prin,
                       H. and Principe, R. and Pugnat, T. and Qiang, J. and
                       Quaranta, E. and Rafique, H. and Rakhno, I. and Duarte, D.
                       Ramos and Ratti, A. and Ravaioli, E. and Raymond, M. and
                       Redaelli, S. and Renaglia, T. and Ricci, D. and Riddone, G.
                       and Rifflet, J. and Rigutto, E. and Rijoff, T. and
                       Rinaldesi, R. and Riu Martinez, O. and Rivkin, L. and
                       Rodriguez Mateos, F. and Roesler, S. and Romera Ramirez, I.
                       and Rossi, A. and Rossi, L. and Rude, V. and Rumolo, G. and
                       Rutkovksi, J. and Sabate Gilarte, M. and Sabbi, G. and
                       Sahner, T. and Salemme, R. and Salvant, B. and Galan, F.
                       Sanchez and Santamaria Garcia, A. and Santillana, I. and
                       Santini, C. and Santos, O. and Diaz, P. Santos and Sasaki,
                       K. and Savary, F. and Sbrizzi, A. and Schaumann, M. and
                       Scheuerlein, C. and Schmalzle, J. and Schmickler, H. and
                       Schmidt, R. and Schoerling, D. and Segreti, M. and Serluca,
                       M. and Serrano, J. and Sestak, J. and Shaposhnikova, E. and
                       Shatilov, D. and Siemko, A. and Sisti, M. and Sitko, M. and
                       Skarita, J. and Skordis, E. and Skoufaris, K. and Skripka,
                       G. and Smekens, D. and Sobiech, Z. and Sosin, M. and
                       Sorbio, M. and Soubelet, F. and Spataro, B. and Spiezia, G.
                       and Stancari, G. and Staterao, M. and Steckert, J. and
                       Steele, G. and Sterbini, G. and Struik, M. and Sugano, M.
                       and Szeberenyi, A. and Taborelli, M. and Tambasco, C. and
                       Rego, R. Tavares and Tavian, L. and Teissandier, B. and
                       Templeton, N. and Therasse, M. and Thiesen, H. and Thomas,
                       E. and Toader, A. and Todesco, E. and Tomás, R. and Toral,
                       F. and Torres-Sanchez, R. and Trad, G. and Triantafyllou,
                       N. and Tropin, I. and Tsinganis, A. and Tuckamantel, J. and
                       Uythoven, J. and Valishev, A. and Van Der Veken, F. and Van
                       Weelderen, R. and Vande Craen, A. and Vazquez De Prada, B.
                       and Velotti, F. and Verdu Andres, S. and Verweij, A. and
                       Shetty, N. Vittal and Vlachoudis, V. and Volpini, G. and
                       Wagner, U. and Wanderer, P. and Wang, M. and Wang, X. and
                       Wanzenberg, R. and Wegscheider, A. and Weisz, S. and
                       Welsch, C. and Wendt, M. and Wenninger, J. and Weterings,
                       W. and White, S. and Widuch, K. and Will, A. and Willering,
                       G. and Wollmann, D. and Wolski, A. and Wozniak, J. and Wu,
                       Q. and Xiao, B. and Xiao, L. and Xu, Q. and Yakovlev, Y.
                       and Yammine, S. and Yang, Y. and Yu, M. and Zacharov, I.
                       and Zagorodnova, O. and Zannini, C. and Zanoni, C. and
                       Zerlauth, M. and Zimmermann, F. and Zlobin, A. and Zobov,
                       M. and Zurbano Fernandez, I.",
      title         = "{High-Luminosity Large Hadron Collider (HL-LHC): Technical
                       design report}",
      publisher     = "CERN",
      address       = "Geneva",
      series        = "CERN Yellow Reports: Monographs",
      year          = "2020",
      url           = "https://cds.cern.ch/record/2749422",
      doi           = "10.23731/CYRM-2020-0010",
}

@article{BENEDIKT2018200,
title = {Proton colliders at the energy frontier},
journal = {Nuclear Instruments and Methods in Physics Research Section A: Accelerators, Spectrometers, Detectors and Associated Equipment},
volume = {907},
pages = {200-208},
year = {2018},
note = {Advances in Instrumentation and Experimental Methods (Special Issue in Honour of Kai Siegbahn)},
issn = {0168-9002},
doi = {https://doi.org/10.1016/j.nima.2018.03.021},
url = {https://www.sciencedirect.com/science/article/pii/S0168900218303577},
author = {Michael Benedikt and Frank Zimmermann}
}

@article{ArkaniHamed:2016,
  title={Physics opportunities of a 100 TeV proton--proton collider},
  author={Arkani-Hamed, Nima and Han, Tao and Mangano, Michelangelo and Wang, Lian-Tao},
  journal={Physics Reports},
  volume={652},
  pages={1--49},
  year={2016},
  publisher={Elsevier}
}

@article{Hoecker:2007,
  title={TMVA-toolkit for multivariate data analysis},
  author={Hoecker, Andreas and Speckmayer, Peter and Stelzer, Joerg and Therhaag, Jan and von Toerne, Eckhard and Voss, Helge and Backes, M and Carli, T and Cohen, O and Christov, A and others},
  journal={arXiv preprint physics/0703039},
  year={2007}
}

@misc{Brun:2019,
  author     = "Brun, Rene and Rademakers, Fons and Canal, Philippe and Naumann, Axel and Couet, Olivier and Moneta, Lorenzo and Vassilev, Vassil and Linev, Sergey and Piparo, Danilo and GANIS, Gerardo and Bellenot, Bertrand and Guiraud, Enrico and Amadio, Guilherme and wverkerke and Mato, Pere and TimurP and Tadel, Matevž and wlav and Tejedor, Enric and Isemann, Raphael",
  title      = "{root-project/root: v6.18/02 (v6-18-02)}",
  repository = "{Zenodo}",
  year       = "2019",
  doi        = "10.5281/zenodo.3895860"
}

@article{Pedregosa:2011,
  title={Scikit-learn: Machine learning in Python},
  author={Pedregosa, Fabian and Varoquaux, Ga{\"e}l and Gramfort, Alexandre and Michel, Vincent and Thirion, Bertrand and Grisel, Olivier and Blondel, Mathieu and Prettenhofer, Peter and Weiss, Ron and Dubourg, Vincent and others},
  journal={the Journal of machine Learning research},
  volume={12},
  pages={2825--2830},
  year={2011},
  publisher={JMLR. org}
}

@article{Paszke:2019,
  title={Pytorch: An imperative style, high-performance deep learning library},
  author={Paszke, A},
  journal={arXiv preprint arXiv:1912.01703},
  year={2019}
}

@misc{Abadi:2015,
title={ {TensorFlow}: Large-Scale Machine Learning on Heterogeneous Systems},
url={https://www.tensorflow.org/},
note={Software available from tensorflow.org},
author={
    Mart\'{i}n~Abadi and
    Ashish~Agarwal and
    Paul~Barham and
    Eugene~Brevdo and
    Zhifeng~Chen and
    Craig~Citro and
    Greg~S.~Corrado and
    Andy~Davis and
    Jeffrey~Dean and
    Matthieu~Devin and
    Sanjay~Ghemawat and
    Ian~Goodfellow and
    Andrew~Harp and
    Geoffrey~Irving and
    Michael~Isard and
    Yangqing Jia and
    Rafal~Jozefowicz and
    Lukasz~Kaiser and
    Manjunath~Kudlur and
    Josh~Levenberg and
    Dandelion~Man\'{e} and
    Rajat~Monga and
    Sherry~Moore and
    Derek~Murray and
    Chris~Olah and
    Mike~Schuster and
    Jonathon~Shlens and
    Benoit~Steiner and
    Ilya~Sutskever and
    Kunal~Talwar and
    Paul~Tucker and
    Vincent~Vanhoucke and
    Vijay~Vasudevan and
    Fernanda~Vi\'{e}gas and
    Oriol~Vinyals and
    Pete~Warden and
    Martin~Wattenberg and
    Martin~Wicke and
    Yuan~Yu and
    Xiaoqiang~Zheng},
  year={2015},
}

@misc{AlRfou:2016,
  title={Theano: A python framework for fast computation of mathematical expressions. arXiv e-prints, arXiv-1605},
  author={Al-Rfou, Rami and Alain, Guillaume and Almahairi, Amjad and Angermueller, Christof and Bahdanau, Dzmitry and Ballas, Nicolas and Bastien, Fr{\'e}d{\'e}ric and Bayer, Justin and Belikov, Anatoly and Belopolsky, Alexander and others},
  year={2016}
}

@inproceedings{SeideAgarwal:2016,
  title={CNTK: Microsoft's open-source deep-learning toolkit},
  author={Seide, Frank and Agarwal, Amit},
  booktitle={Proceedings of the 22nd ACM SIGKDD international conference on knowledge discovery and data mining},
  pages={2135--2135},
  year={2016}
}

@misc{Chollet:2021,
  title={Keras},
  author={Chollet, Fran\c{c}ois and others},
  year={2015},
  howpublished={\url{https://keras.io}},
}

@misc{FeickertNachman:2021,
      title={A Living Review of Machine Learning for Particle Physics}, 
      author={Matthew Feickert and Benjamin Nachman},
      year={2021},
      eprint={2102.02770},
      archivePrefix={arXiv},
      primaryClass={hep-ph},
      url={https://arxiv.org/abs/2102.02770}, 
}

@misc{GrossoLetizia:2024,
      title={Multiple testing for signal-agnostic searches of new physics with machine learning}, 
      author={Gaia Grosso and Marco Letizia},
      year={2024},
      eprint={2408.12296},
      archivePrefix={arXiv},
      primaryClass={hep-ph},
      url={https://arxiv.org/abs/2408.12296}, 
}

@misc{OpenAIo3:2025,
  author     = "OpenAI",
  title      = "{Introducing OpenAI o3 and o4-mini}",
  repository = "{OpenAI}",
  doi        = "",
  year       = "2025"
}

@article{LiuMoretti:2021,
doi = {10.1088/1674-1137/abe0c0},
url = {https://dx.doi.org/10.1088/1674-1137/abe0c0},
year = {2021},
month = {apr},
publisher = {Chinese Physical Society and the Institute of High Energy Physics of the Chinese Academy of Sciences and the Institute of Modern Physics of the Chinese Academy of Sciences and IOP Publishing Ltd},
volume = {45},
number = {4},
pages = {043110},
author = {Liu, Yao-Bei and Moretti, Stefano},
title = {Probing tqZ anomalous couplings in the trilepton signal at the HL-LHC, HE-LHC, and FCC-hh *},
journal = {Chinese Physics C}
}

@misc{OpenAI_CP:2025,
      title={Competitive Programming with Large Reasoning Models}, 
      author={OpenAI and : and Ahmed El-Kishky and Alexander Wei and Andre Saraiva and Borys Minaiev and Daniel Selsam and David Dohan and Francis Song and Hunter Lightman and Ignasi Clavera and Jakub Pachocki and Jerry Tworek and Lorenz Kuhn and Lukasz Kaiser and Mark Chen and Max Schwarzer and Mostafa Rohaninejad and Nat McAleese and o3 contributors and Oleg Mürk and Rhythm Garg and Rui Shu and Szymon Sidor and Vineet Kosaraju and Wenda Zhou},
      year={2025},
      eprint={2502.06807},
      archivePrefix={arXiv},
      primaryClass={cs.LG},
      url={https://arxiv.org/abs/2502.06807}, 
}

@article{aguilar2009,
  title={A Minimal set of top anomalous couplings},
  author={Aguilar-Saavedra, Juan Antonio},
  journal={Nuclear Physics B},
  volume={812},
  number={1-2},
  pages={181--204},
  year={2009},
  publisher={Elsevier}
}

@article{li1991qcd,
  title={QCD corrections to t→ W++ b},
  author={Li, Chong Sheng and Oakes, Robert J and Yuan, Tzu Chiang},
  journal={Physical Review D},
  volume={43},
  number={11},
  pages={3759},
  year={1991},
  publisher={APS}
}

@article{zhang2009next,
  title={Next-to-leading-order QCD corrections to the top-quark decay via model-independent flavor-changing neutral-current couplings},
  author={Zhang, Jia Jun and Li, Chong Sheng and Gao, Jun and Zhang, Hao and Li, Zhao and Yuan, C-P and Yuan, Tzu-Chiang},
  journal={Physical review letters},
  volume={102},
  number={7},
  pages={072001},
  year={2009},
  publisher={APS}
}

@article{drobnak2010flavor,
  title={Flavor Changing Neutral Coupling Mediated Radiative Top Quark Decays at Next-to-Leading Order in QCD},
  author={Drobnak, Jure and Fajfer, Svjetlana and Kamenik, Jernej F},
  journal={Physical review letters},
  volume={104},
  number={25},
  pages={252001},
  year={2010},
  publisher={APS}
}

@article{alloul2014feynrules,
  title={FeynRules 2.0—A complete toolbox for tree-level phenomenology},
  author={Alloul, Adam and Christensen, Neil D and Degrande, C{\'e}line and Duhr, Claude and Fuks, Benjamin},
  journal={Computer Physics Communications},
  volume={185},
  number={8},
  pages={2250--2300},
  year={2014},
  publisher={Elsevier}
}

@article{degrande2012ufo,
  title={UFO--the universal FeynRules output},
  author={Degrande, Celine and Duhr, Claude and Fuks, Benjamin and Grellscheid, David and Mattelaer, Olivier and Reiter, Thomas},
  journal={Computer Physics Communications},
  volume={183},
  number={6},
  pages={1201--1214},
  year={2012},
  publisher={Elsevier}
}

@article{alwall2014automated,
  title={The automated computation of tree-level and next-to-leading order differential cross sections, and their matching to parton shower simulations},
  author={Alwall, Johan and Frederix, Rikkert and Frixione, Stefano and Hirschi, Valentin and Maltoni, Fabio and Mattelaer, Olivier and Shao, H-S and Stelzer, Tim and Torrielli, Paolo and Zaro, Marco},
  journal={Journal of High Energy Physics},
  volume={2014},
  number={7},
  pages={1--157},
  year={2014},
  publisher={Springer}
}

@article{ball2014nnpdf,
  title={NNPDF},
  author={Ball, Richard D and Carrazza, Stefano and Del Debbio, Luigi and Forte, Stefano and Kassabov, Zahari and Rojo, Juan and Slade, Emma and Ubiali, Maria},
  journal={Parton distributions from high-precision collider data,” Eur. Phys. J. C},
  volume={77},
  year={2014}
}

@article{bierlich2022comprehensive,
  title={A comprehensive guide to the physics and usage of PYTHIA 8.3},
  author={Bierlich, Christian and Chakraborty, Smita and Desai, Nishita and Gellersen, Leif and Helenius, Ilkka and Ilten, Philip and L{\"o}nnblad, Leif and Mrenna, Stephen and Prestel, Stefan and Preuss, Christian Tobias and others},
  journal={SciPost Physics Codebases},
  pages={008},
  year={2022}
}

@article{cacciari2012fastjet,
  title={FastJet user manual: (for version 3.0. 2)},
  author={Cacciari, Matteo and Salam, Gavin P and Soyez, Gregory},
  journal={The European Physical Journal C},
  volume={72},
  pages={1--54},
  year={2012},
  publisher={Springer}
}

@article{cacciari2008anti,
  title={The anti-kt jet clustering algorithm},
  author={Cacciari, Matteo and Salam, Gavin P and Soyez, Gregory},
  journal={Journal of High Energy Physics},
  volume={2008},
  number={04},
  pages={063},
  year={2008},
  publisher={IOP Publishing}
}

@article{de2014delphes,
  title={DELPHES 3: a modular framework for fast simulation of a generic collider experiment},
  author={De Favereau, J and Delaere, Christophe and Demin, Pavel and Giammanco, Andrea and Lemaitre, Vincent and Mertens, Alexandre and Selvaggi, Michele},
  journal={Journal of High Energy Physics},
  volume={2014},
  number={2},
  pages={1--26},
  year={2014},
  publisher={Springer}
}

@article{conte2013madanalysis,
  title={MadAnalysis 5, a user-friendly framework for collider phenomenology},
  author={Conte, Eric and Fuks, Benjamin and Serret, Guillaume},
  journal={Computer Physics Communications},
  volume={184},
  number={1},
  pages={222--256},
  year={2013},
  publisher={Elsevier}
}

@article{li2011next,
  title={Next-to-leading order QCD corrections to t Z associated production via the flavor-changing neutral-current couplings at hadron colliders},
  author={Li, Bo Hua and Zhang, Yue and Li, Chong Sheng and Gao, Jun and Zhu, Hua Xing},
  journal={Physical Review D—Particles, Fields, Gravitation, and Cosmology},
  volume={83},
  number={11},
  pages={114049},
  year={2011},
  publisher={APS}
}

@article{zhang2011effective,
  title={Effective-field-theory approach to top-quark production and decay},
  author={Zhang, Cen and Willenbrock, Scott},
  journal={Physical Review D—Particles, Fields, Gravitation, and Cosmology},
  volume={83},
  number={3},
  pages={034006},
  year={2011},
  publisher={APS}
}

@article{degrande2015automatic,
  title={Automatic computations at next-to-leading order in QCD for top-quark flavor-changing neutral processes},
  author={Degrande, Celine and Maltoni, Fabio and Wang, Jian and Zhang, Cen},
  journal={Physical Review D},
  volume={91},
  number={3},
  pages={034024},
  year={2015},
  publisher={APS}
}

@article{lazopoulos2008next,
  title={Next-to-leading order QCD corrections to tt Z production at the LHC},
  author={Lazopoulos, Achilleas and McElmurry, Thomas and Melnikov, Kirill and Petriello, Frank},
  journal={Physics Letters B},
  volume={666},
  number={1},
  pages={62--65},
  year={2008},
  publisher={Elsevier}
}

@article{kardos2012top,
  title={Top quark pair production in association with a Z-boson at next-to-leading-order accuracy},
  author={Kardos, Adam and Trocsanyi, Zoltan and Papadopoulos, CG},
  journal={Physical Review D—Particles, Fields, Gravitation, and Cosmology},
  volume={85},
  number={5},
  pages={054015},
  year={2012},
  publisher={APS}
}

@article{czakon2013total,
  title={Total top-quark pair-production cross section at hadron colliders through O ($\alpha$ S 4)},
  author={Czakon, Micha{\l} and Fiedler, Paul and Mitov, Alexander},
  journal={Physical Review Letters},
  volume={110},
  number={25},
  pages={252004},
  year={2013},
  publisher={APS}
}

@misc{https://doi.org/10.23731/cyrm-2017-003,
  doi = {10.23731/CYRM-2017-003},
  
  url = {https://e-publishing.cern.ch/index.php/CYRM/issue/view/35},
  
  author = {{CERN}
},
  
  language = {en},
  
  title = {CERN Yellow Reports: Monographs, Vol 3 (2017): Physics at the FCC-hh, a 100 TeV pp collider},
  
  publisher = {CERN},
  
  year = {2017},
  
  copyright = {This work is licensed under a Creative Commons Attribution 4.0 International License.}
}

@article{campanario2010nlo,
  title={NLO QCD corrections to WZ+ jet production with leptonic decays},
  author={Campanario, Francisco and Englert, Christoph and Kallweit, Stefan and Spannowsky, Michael and Zeppenfeld, Dieter},
  journal={Journal of High Energy Physics},
  volume={2010},
  number={7},
  pages={1--15},
  year={2010},
  publisher={Springer}
}

@article{campbell2012t,
  title = {$t\bar{t}W^{\pm}$ production and decay at NLO},
  author = {Campbell, John M. and Ellis, R. Keith},
  journal = {Journal of High Energy Physics},
  volume = {2012},
  number = {7},
  pages = {1--12},
  year = {2012},
  publisher = {Springer}
}

@article{frederix2018large,
  title = {Large NLO corrections in $t\bar{t}W^{\pm}$ and $t\bar{t}t\bar{t}$ hadroproduction from supposedly subleading EW contributions},
  author = {Frederix, Rikkert and Pagani, Davide and Zaro, Marco},
  journal = {Journal of High Energy Physics},
  volume = {2018},
  number = {2},
  pages = {1--39},
  year = {2018},
  publisher = {Springer}
}

@article{frixione2015electroweak,
  title={Electroweak and QCD corrections to top-pair hadroproduction in association with heavy bosons},
  author={Frixione, Stefano and Hirschi, V and Pagani, Davide and Shao, H-S and Zaro, Marco},
  journal={Journal of High Energy Physics},
  volume={2015},
  number={6},
  pages={1--28},
  year={2015},
  publisher={Springer}
}

@misc{fowlie2016lhcoreadernewcodereading,
  author       = {Fowlie, Andrew},
  title        = {LHCO\_reader: A new code for reading and analyzing detector-level events stored in LHCO format},
  year         = {2016},
  eprint       = {1510.07319},
  archivePrefix = {arXiv},
  primaryClass = {hep-ph},
  note         = {arXiv:1510.07319 [hep-ph]},
  url          = {https://arxiv.org/abs/1510.07319}
}

%% or
%% [B] Manual formatting (see below)
%% (i) We suggest to always provide author, title and journal data or doi:
%% in short all the informations that clearly identify a document.
%% (ii) please avoid comments such as "For a review'', "For some examples",
%% "and references therein" or move them in the text. In general, please leave only references in the bibliography and move all
%% accessory text in footnotes.
%% (iii) Also, please have only one work for each \bibitem.

\newpage

\appendix
\section{Appendix}\label{app:fig}

Below are the o3 model prompts provided which were utilized in this study:\\

Prompt 1:
\begin{verbatim}
I have attached 5 lhco files containing signals and backgrounds.
Signal:
ugtz.lhco
Background:
tt.lhco
ttz.lhco
ttw.lhco
wz.lhco
The processes for signal and backgrounds are given as:
Signal:
p p -> t Z, (t -> w+ b, w+ -> l+ vl~), (Z -> l+ l-)
(focus on the ugtZ processes here)
Backgrounds:
p p -> t t~ Z, (t -> w+ b, w+ -> l+ vl~), (t~ -> w- b~, w- -> l- vl),
(Z -> l+ l-)
p p -> t t~, (t -> w+ b, w+ -> l+ vl~), (t~ -> w- b~, w- -> l- vl)
p p -> t t~ W+, (t -> w+ b, w+ -> l+ vl~), (t~ -> w- b~, w- -> l- vl),
(w+ -> l+ vl~)
p p -> W+ Z, (w+ -> l+ vl~), (Z -> l+ l-)
the format of a typical lhco file is given as (with the first line starting
with #):
# typ eta phi pt jmas ntrk btag had/em dum1 dum2
0 0 0
1 0 5.061 2.658 1.12 0.00 0.0 0.0 0.00 0.0 0.0
2 0 -3.509 -2.564 1.04 0.00 0.0 0.0 0.00 0.0 0.0
3 0 -2.782 0.590 1.02 0.00 0.0 0.0 0.00 0.0 0.0
4 1 -1.440 1.228 72.70 0.00 -1.0 0.0 0.00 0.0 0.0
5 1 -3.358 -1.410 70.78 0.00 -1.0 0.0 0.00 0.0 0.0
6 4 -3.596 2.860 42.47 1.85 2.0 0.0 0.13 0.0 0.0
7 6 0.000 -0.178 13.64 0.00 0.0 0.0 0.00 0.0 0.0
0 1 0
1 0 1.052 1.427 1.62 0.00 0.0 0.0 0.00 0.0 0.0
2 1 0.324 -1.944 40.58 0.00 1.0 0.0 0.00 0.0 0.0
3 2 0.449 1.114 37.05 0.11 1.0 0.0 38.99 0.0 0.0
4 2 -0.684 -2.080 37.03 0.11 -1.0 0.0 38.99 0.0 0.0
5 6 0.000 1.206 49.55 0.00 0.0 0.0 0.00 0.0 0.0
0 2 0
lhco charachter notations:
0 = photon
1 = electron
2 = muon

Ok, let's go over it part by part again. parse through the lhco files provided;
I want you to provide the first cut on:
number and charge of leptons
number of b-quark
PT of lepton, sub-leading leptons,sub-sub-leading leptons b, jets (or any other
subleading jets and so on)  utilizing the info that I have provided with the
aim of maximum background removal!

\end{verbatim}

Prompt 2:
\begin{verbatim}   
In each of the file, parse through and
now apply cut on the Opposite sign same flavor lepton pair that form the Z boson 
in the original signal:
Apply two cuts here:
DeltaR (OSSF pair) (here you can try to devise a cut that removes as much
background as possible Range on invariant mass of Z constructed from this
lepton pair. Here since almost all background have a Z boson, try to limit the 
range to an appropriate degree that takes out most of the other discrepancies 
that may occur since you can't really distinguish this much from the background
Z's.

\end{verbatim}

Prompt 3:
\begin{verbatim}
Nowparse through the .lhco files again and apply a cut on the rapidity of the
same OSSF pair that formed the Z boson in order to eliminate as much background 
as possible while preserving the signal.

\end{verbatim}

Prompt 4:
\begin{verbatim}
Now calculate the transverse masses of W boson and top quark from the files
provided and then produce range cuts on the masses of both with the aim of
elminating the background as much as possible. Since the background signals
also contain t and Z, the purpose of the cut is to be more precise at the peak,
eliminating the noise rather than to delete the backgrounds fully
\end{verbatim}
\section{Appendix}\label{app:fig1}
Below is the graphical comparison of variables between the signal and background events:
\begin{figure}[ht!]
    \centering

    % First row
    \begin{subfigure}[b]{0.32\textwidth}
        \includegraphics[width=\textwidth]{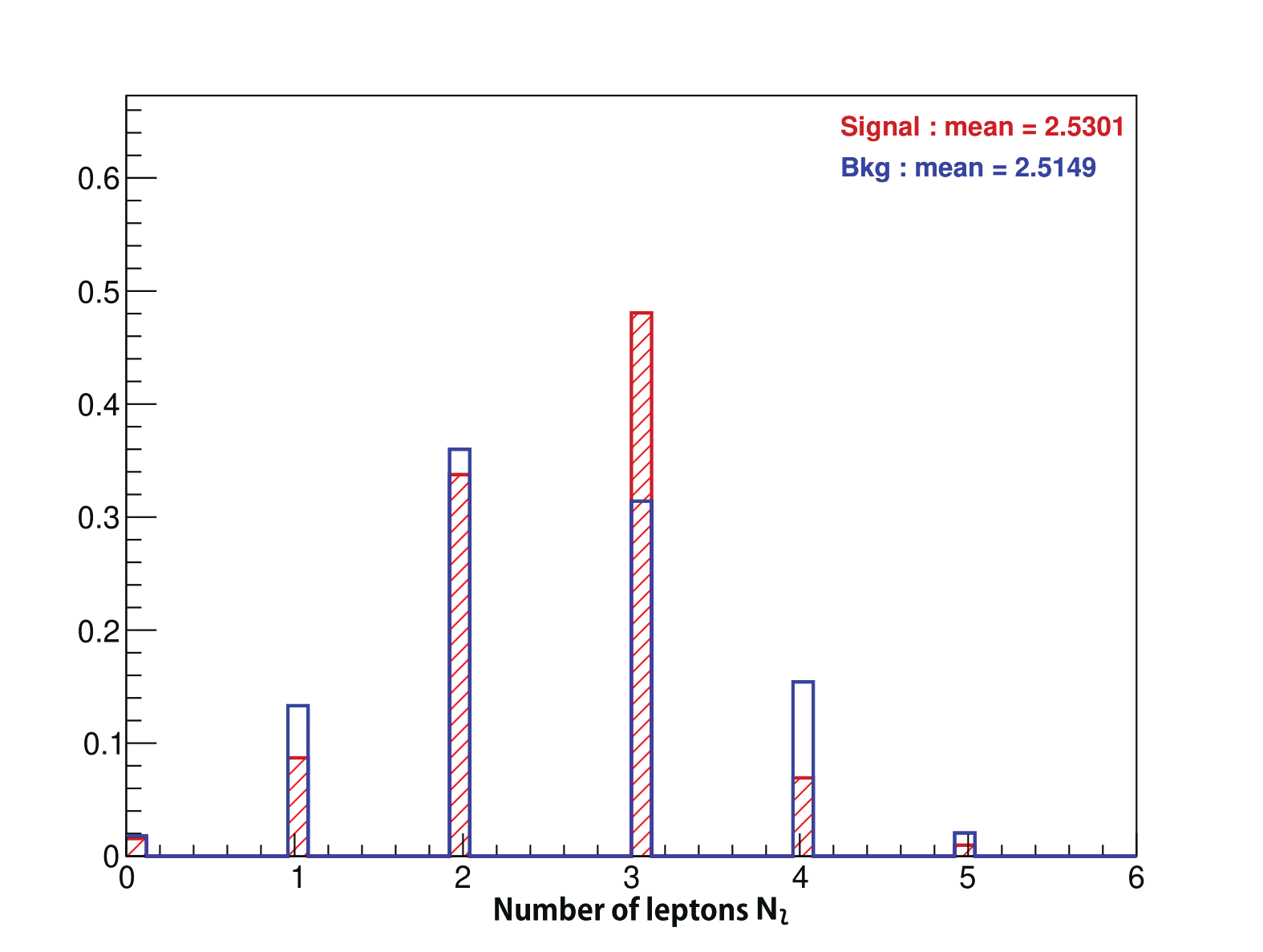}
    \end{subfigure}
    \begin{subfigure}[b]{0.32\textwidth}
        \includegraphics[width=\textwidth]{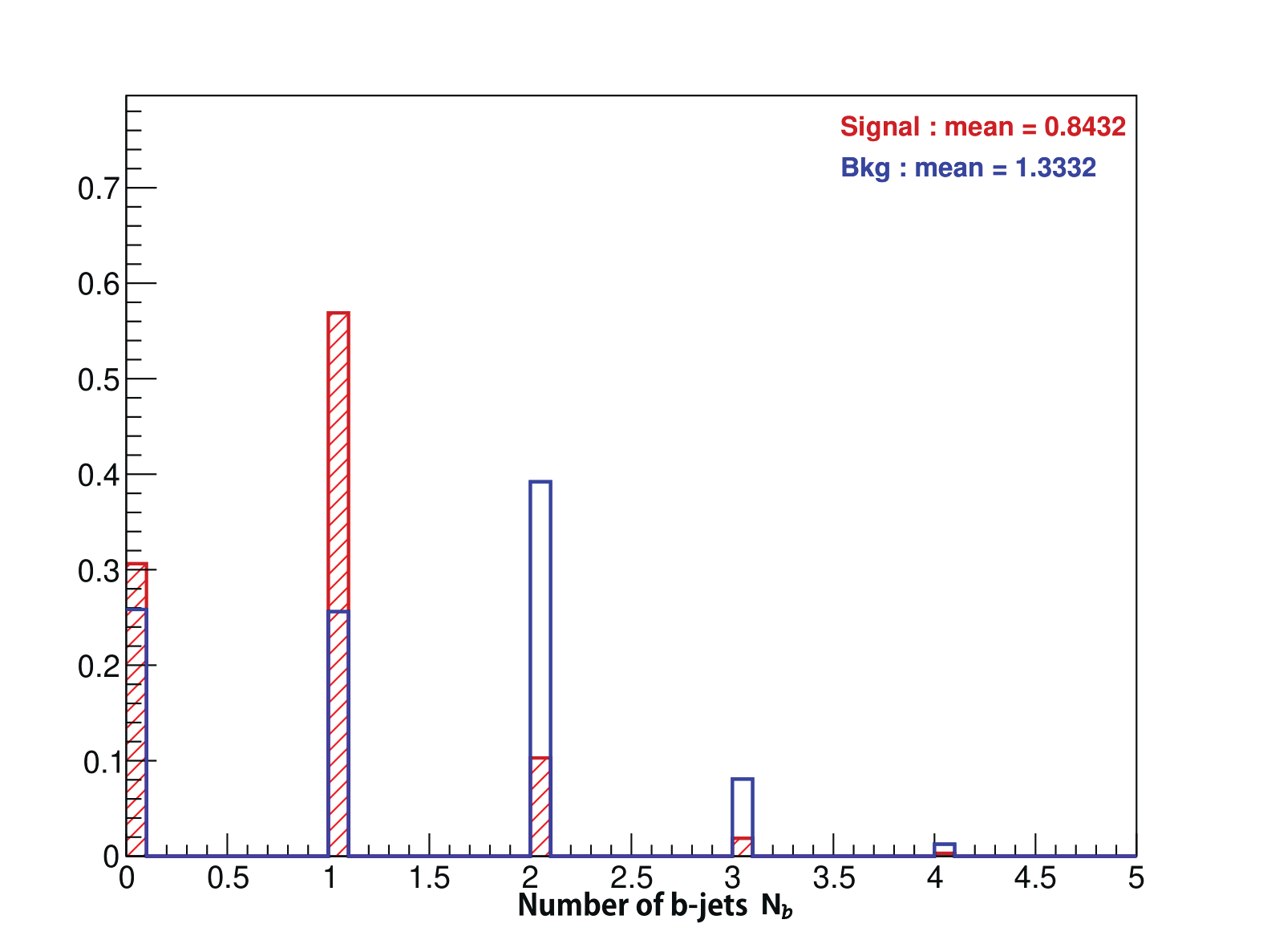}
    \end{subfigure}
    \begin{subfigure}[b]{0.32\textwidth}
        \includegraphics[width=\textwidth]{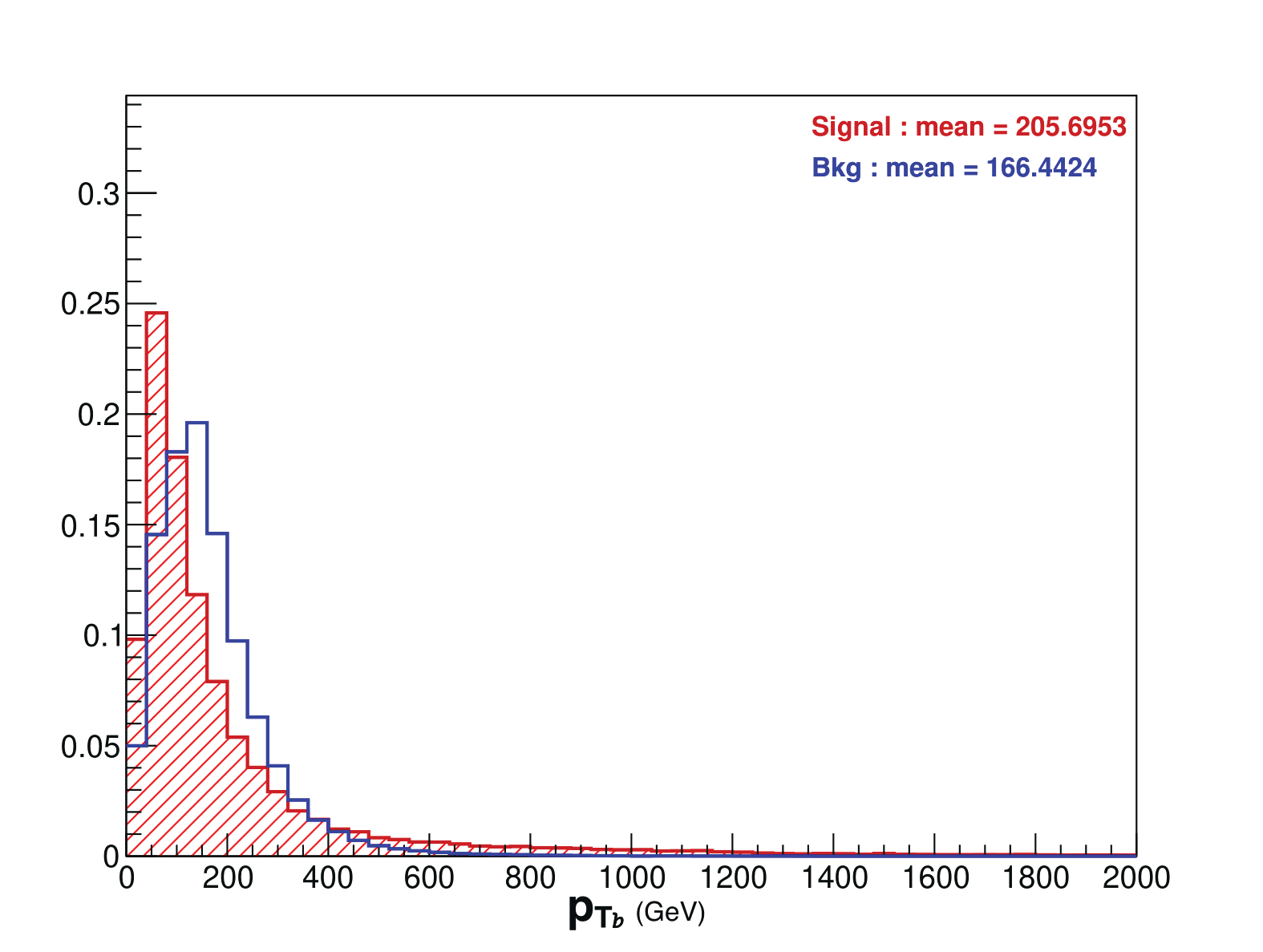}
    \end{subfigure}

    \vspace{0.5em} % optional spacing between rows

    % Second row
    \begin{subfigure}[b]{0.32\textwidth}
        \includegraphics[width=\textwidth]{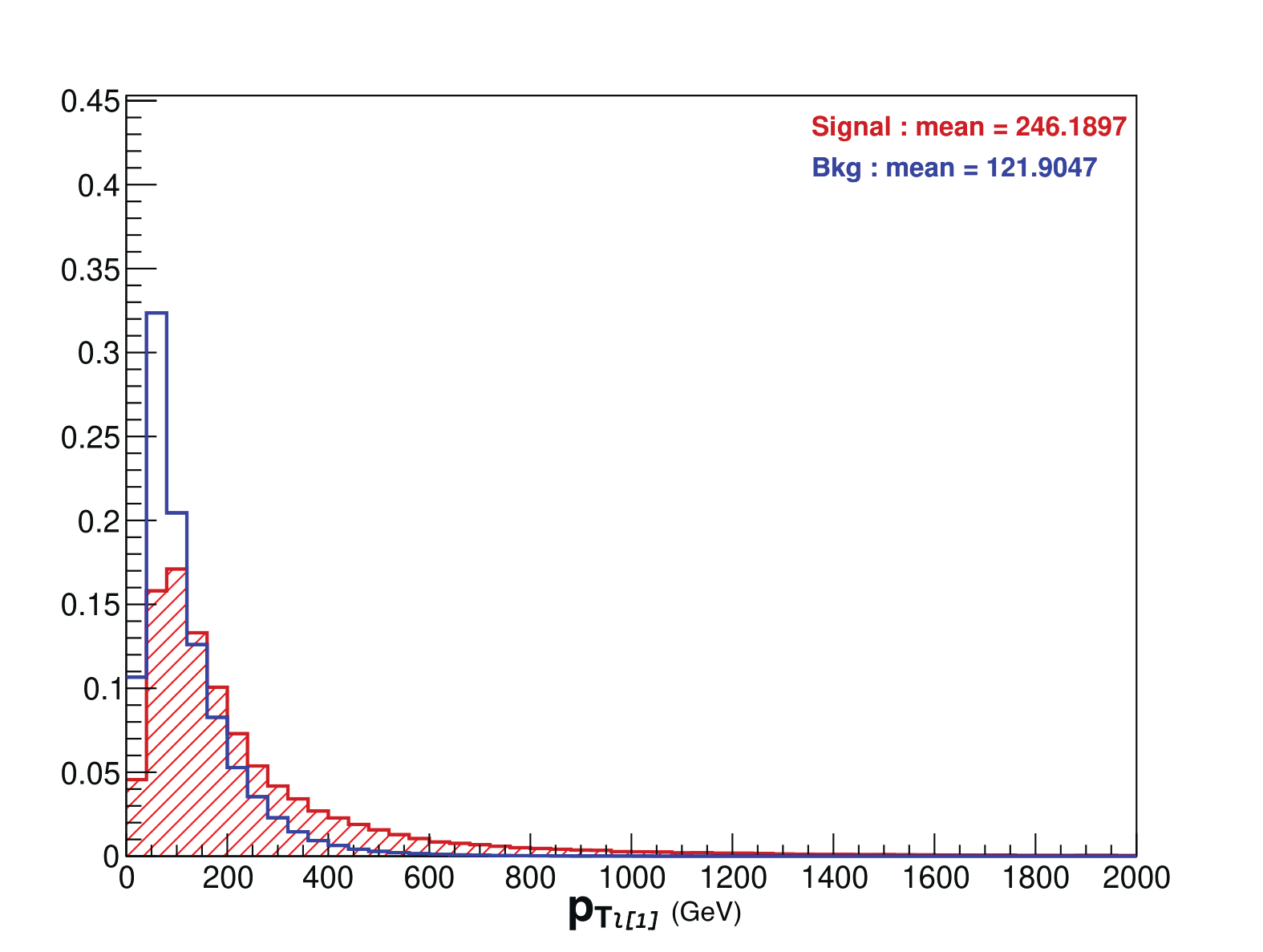}
    \end{subfigure}
    \begin{subfigure}[b]{0.32\textwidth}
        \includegraphics[width=\textwidth]{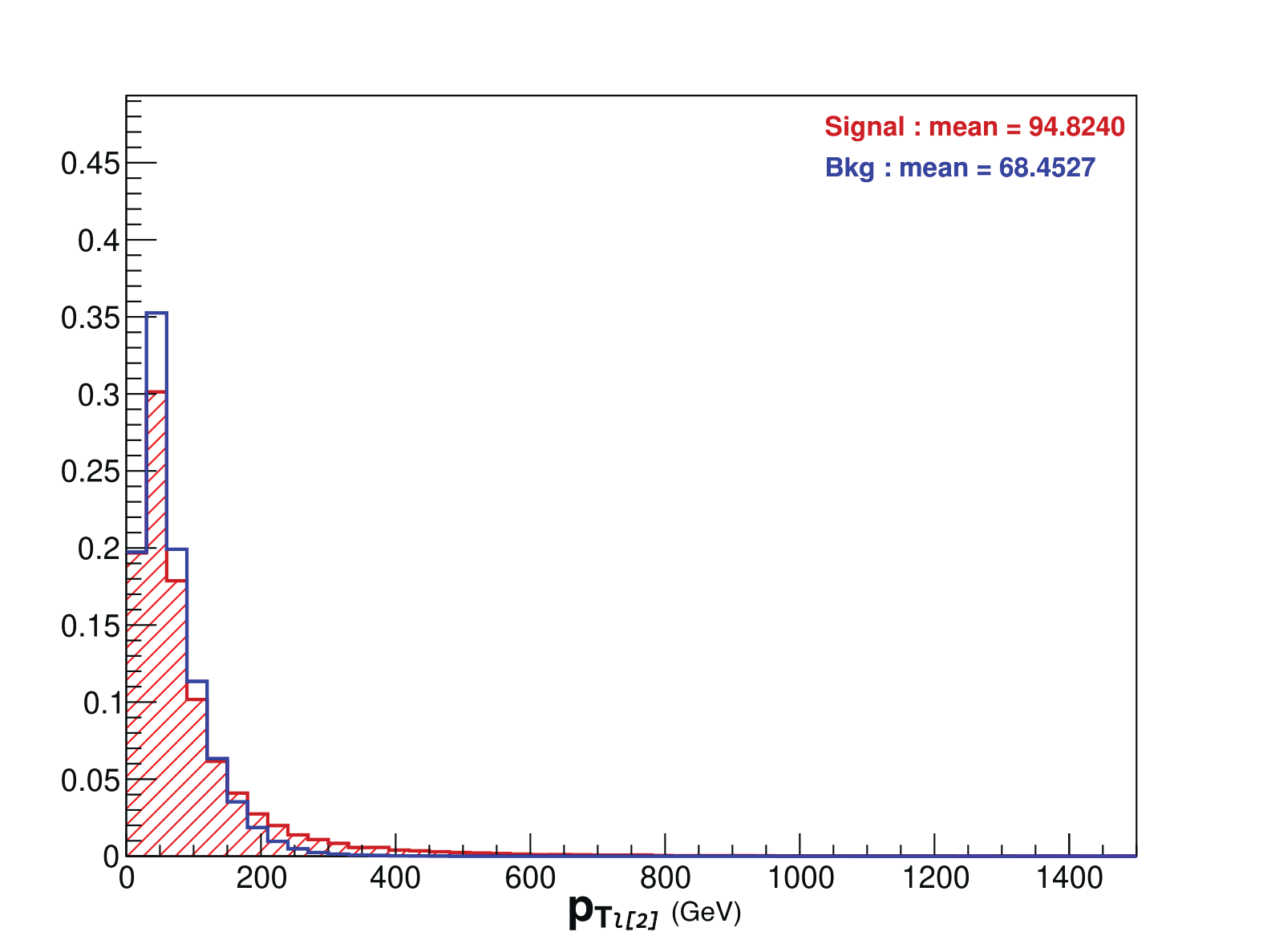}
    \end{subfigure}
    \begin{subfigure}[b]{0.32\textwidth}
        \includegraphics[width=\textwidth]{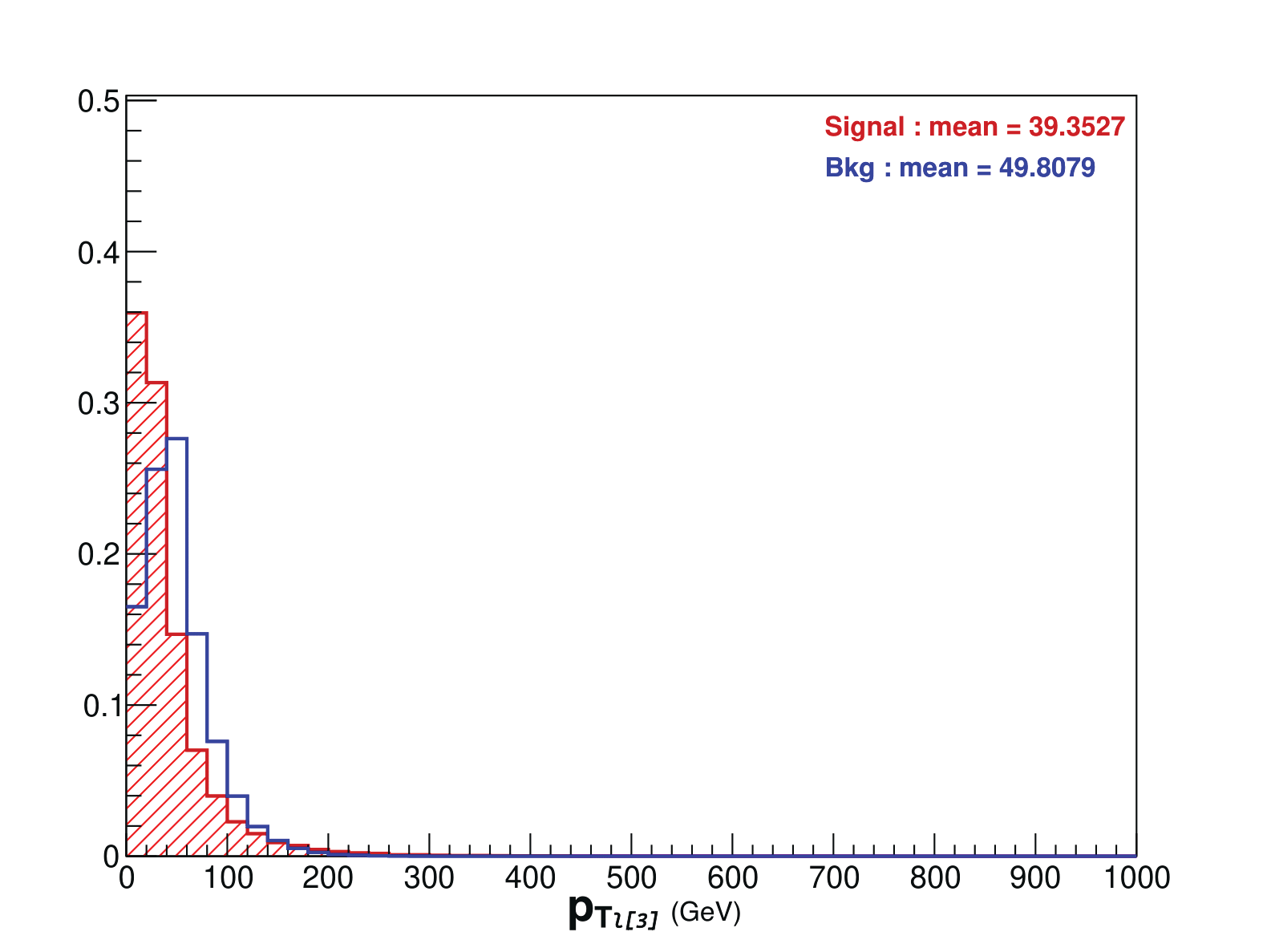}
    \end{subfigure}

    % Third row
    \begin{subfigure}[b]{0.32\textwidth}
        \includegraphics[width=\textwidth]{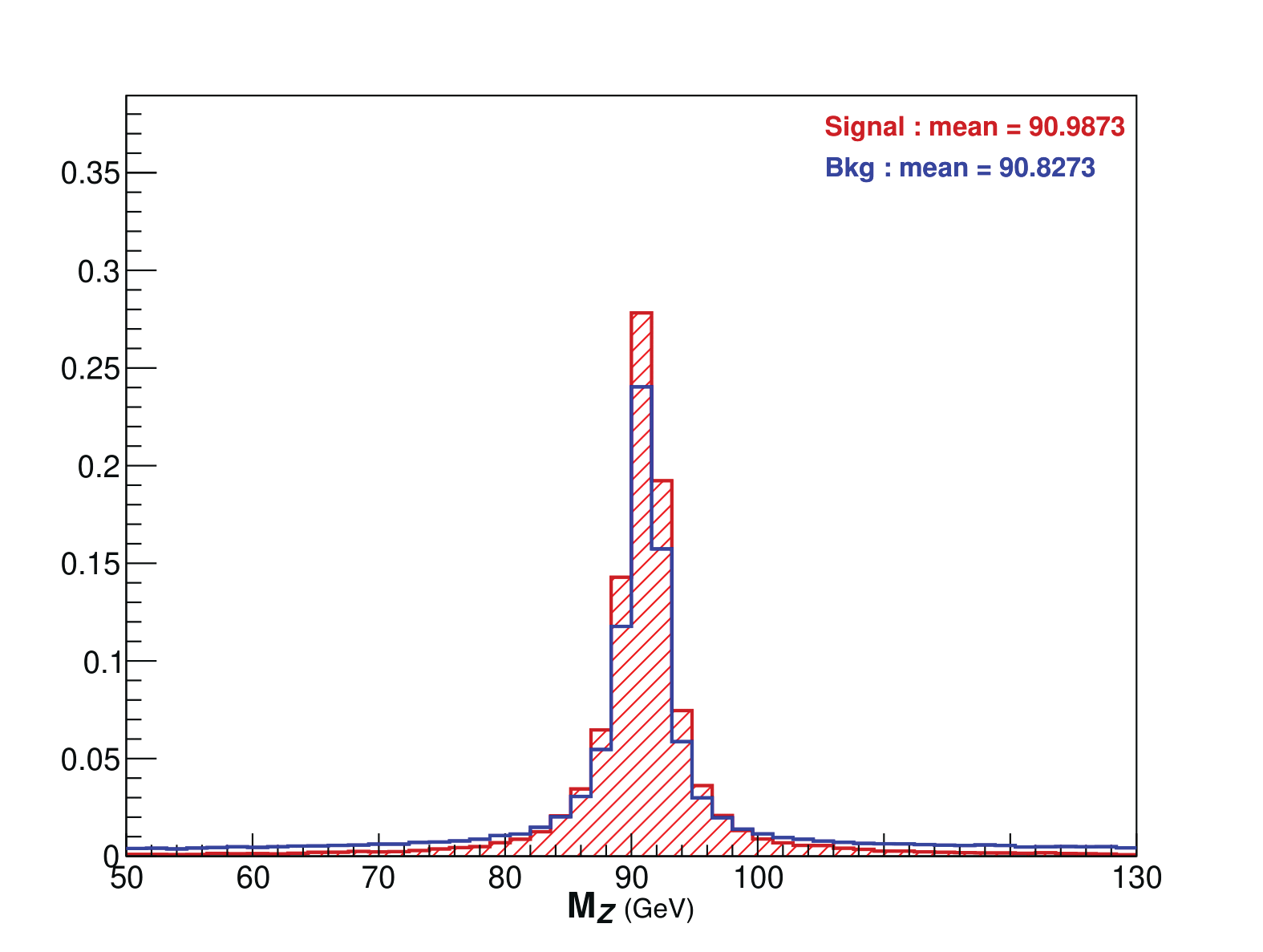}
    \end{subfigure}
    \begin{subfigure}[b]{0.32\textwidth}
        \includegraphics[width=\textwidth]{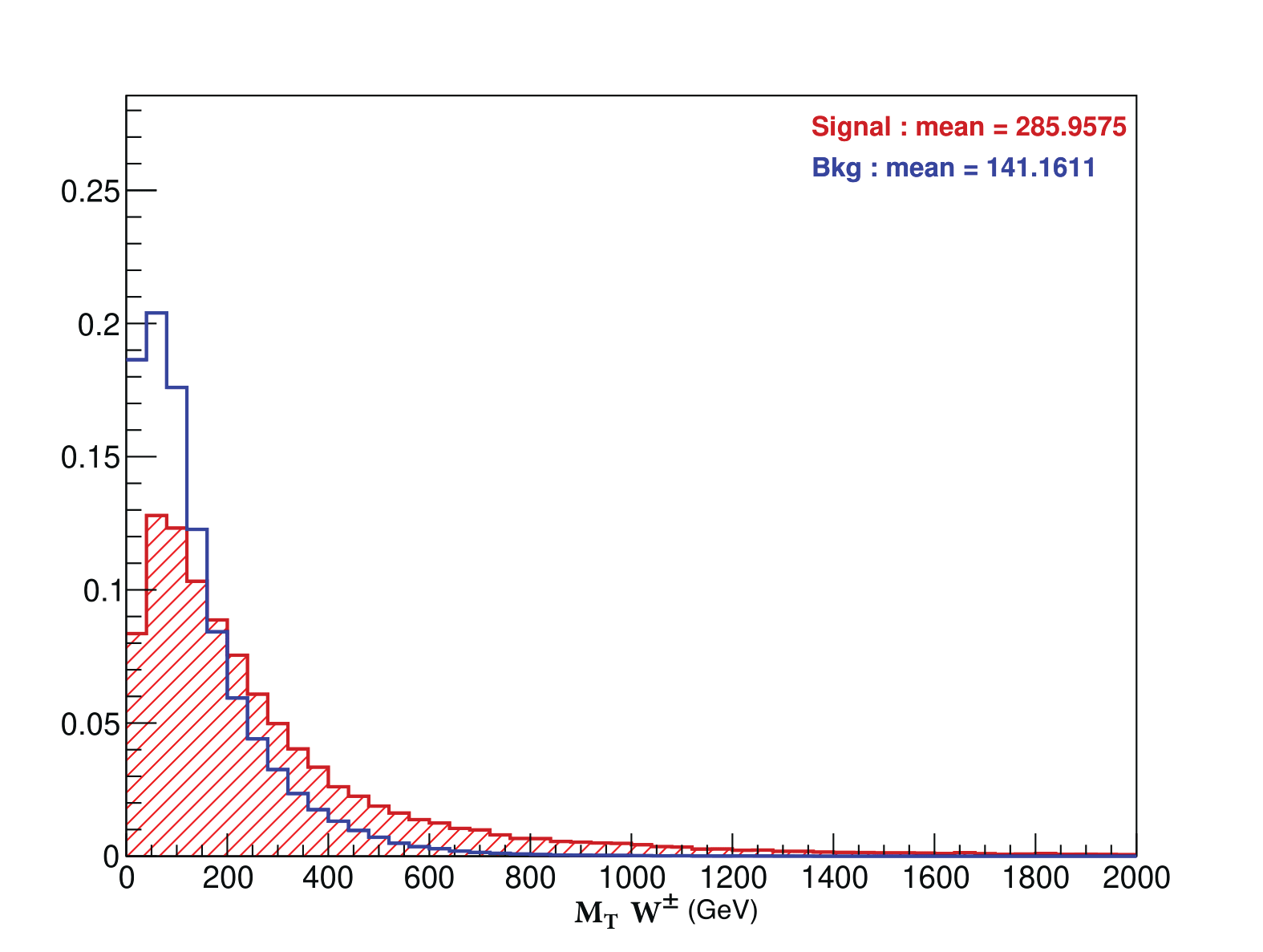}
    \end{subfigure}
    \begin{subfigure}[b]{0.32\textwidth}
        \includegraphics[width=\textwidth]{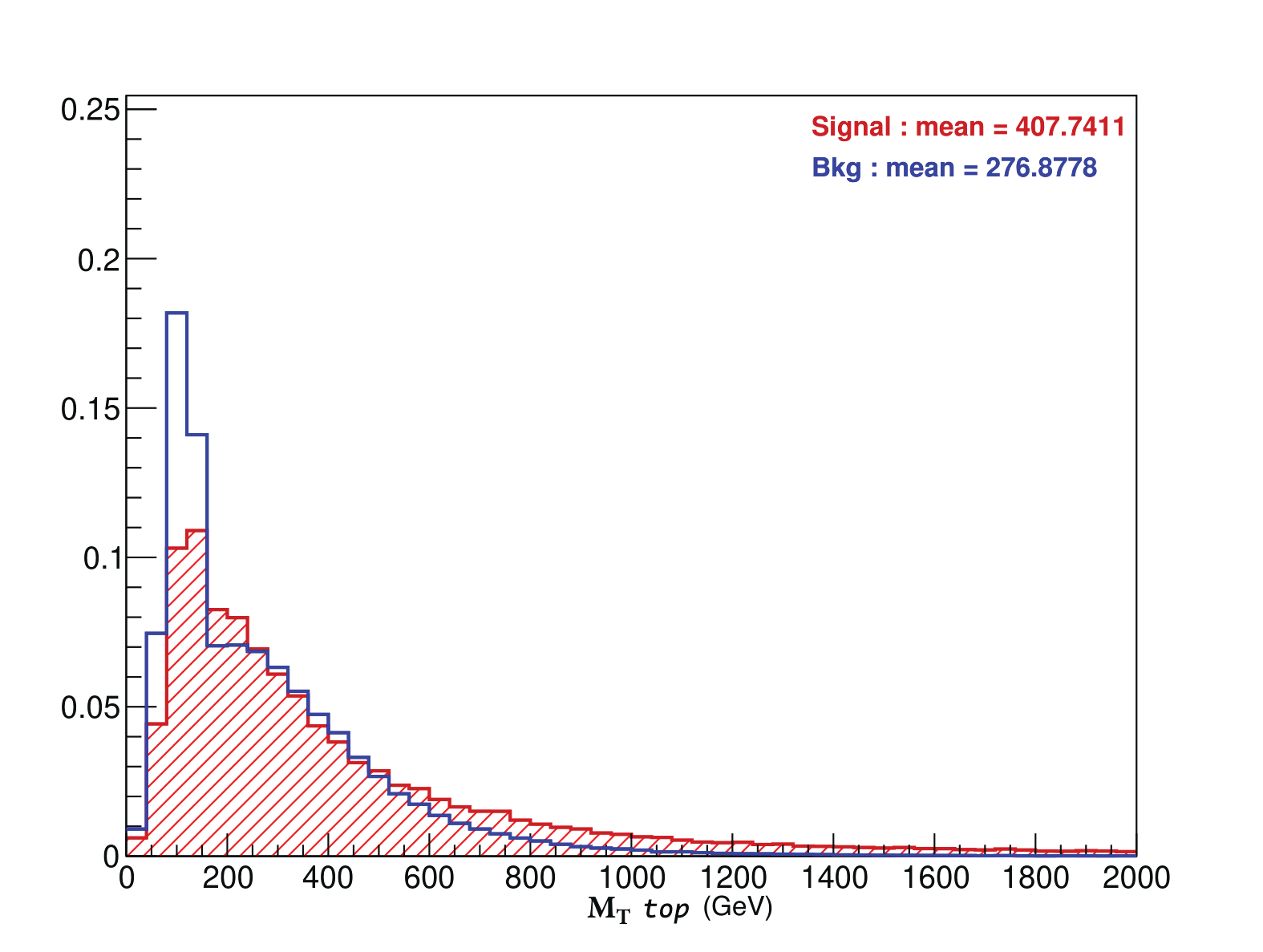}
    \end{subfigure}

        % Third row
    \begin{subfigure}[b]{0.32\textwidth}
        \includegraphics[width=\textwidth]{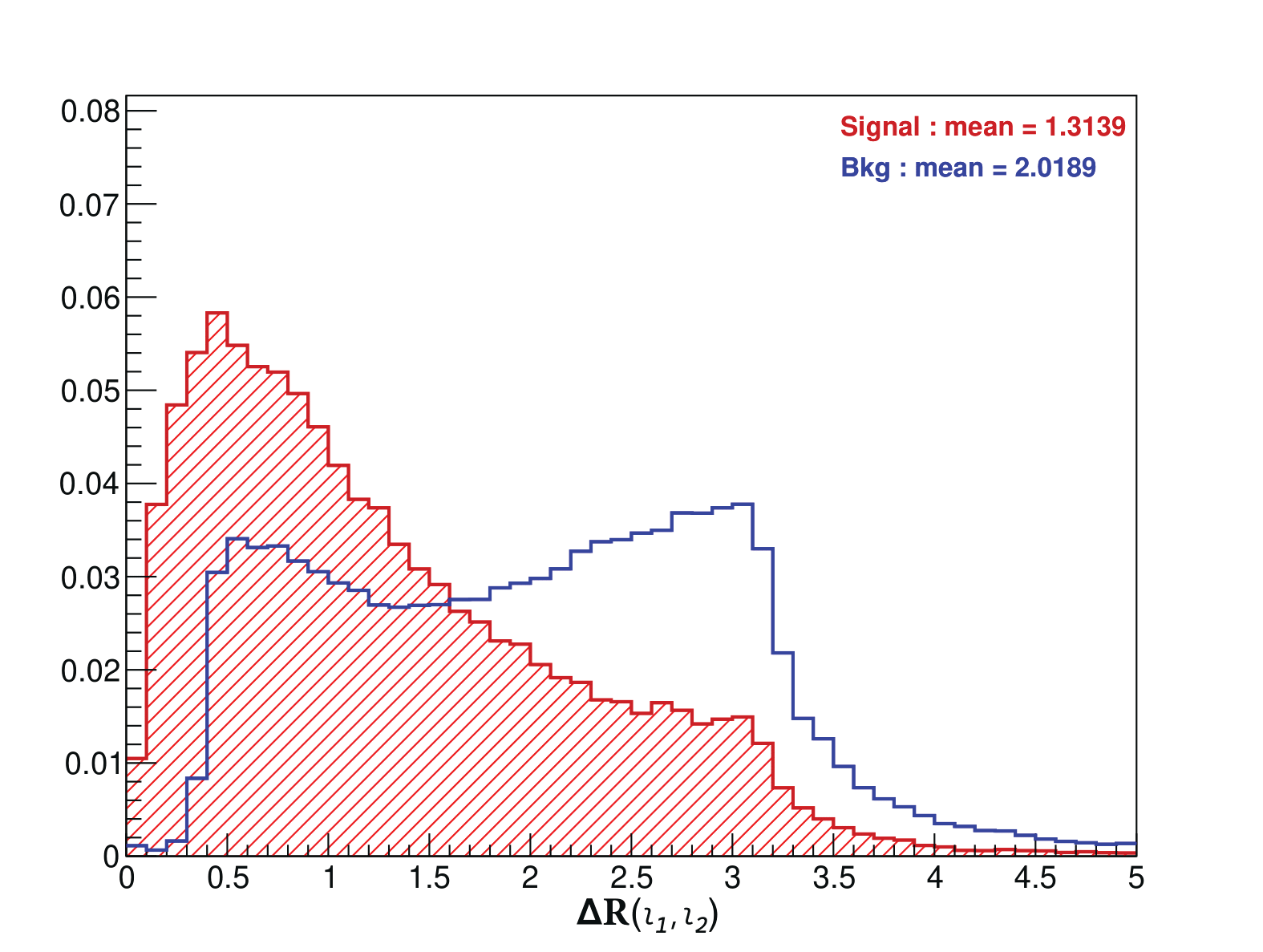}
    \end{subfigure}
    \begin{subfigure}[b]{0.32\textwidth}
        \includegraphics[width=\textwidth]{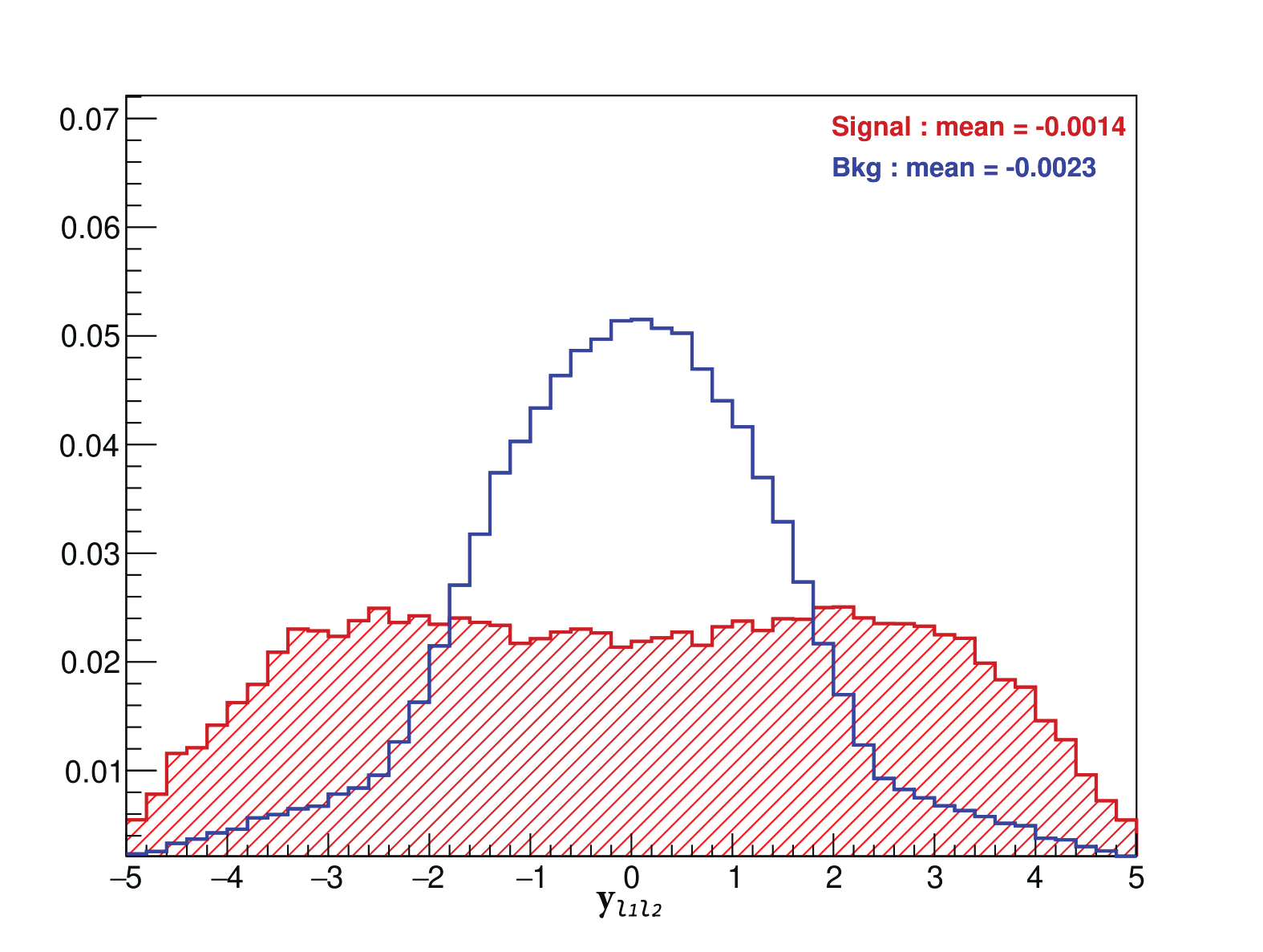}
    \end{subfigure}

\end{figure}

\end{document}